\documentclass[12pt,preprint]{aastex}

\catcode`\@=11
\def\gsim{\,\mathrel{\mathpalette\@versim>\,}}
\def\lsim{\,\mathrel{\mathpalette\@versim<\,}}
\def\@versim#1#2{\lower 2.9truept \vbox{\baselineskip 0pt \lineskip
    0.5truept \ialign{$\m@th#1\hfil##\hfil$\crcr#2\crcr\sim\crcr}}}
\catcode`\@=12  
\newcommand{\az}{a_0}
\newcommand{\gz}{g_0}
\newcommand{\gv}{{\bf g}}

\newcommand{\xv}{{\bf x}}
\newcommand{\de}{{\rm d}}
\newcommand{\rhoz}{\rho_0}
\newcommand{\varrhoz}{\varrho_0}
\newcommand{\phiz}{\phi_0}

\newcommand{\phiu}{\phi_1}
\newcommand{\Mz}{M_0}
\newcommand{\rhoNz}{\rho_{\rm N0}}
\newcommand{\rhoMz}{\rho_{\rm M0}}

\newcommand{\rhoNu}{\rho_{\rm N1}}
\newcommand{\phiN}{\phi_{\rm N}}
\newcommand{\psiu}{\psi_1}
\newcommand{\psid}{\psi_2}
\newcommand{\gvnum}{{\bf g}_{\rm num}}
\newcommand{\phinum}{\phi_{\rm num}}
\newcommand{\gvN}{\gv_{\rm N}}
\newcommand{\gN}{g_{\rm N}}
\newcommand{\hv}{{\bf h}}
\newcommand{\Svu}{{\bf S}_1}
\newcommand{\Sv}{{\bf S}}
\newcommand{\rc}{r_{\rm c}}    
\newcommand{\sthsq}{\sin^2\vartheta}
\newcommand{\cthsq}{\cos^2\vartheta}
\newcommand{\sphsq}{\sin^2\varphi}
\newcommand{\sth}{\sin\vartheta}
\newcommand{\cth}{\cos\vartheta}

\newcommand{\etatilde}{\tilde{\eta}}
\newcommand{\epsilontilde}{\tilde{\epsilon}}

\newcommand{\be}{\begin{equation}}
\newcommand{\ee}{\end{equation}}

\def\xx {{\bf x}}

\def\phinow{\phi^{(n)}}
\def\phinowlm{\phi_{l,m}^{(n)}}
\def\phiplus{\phi^{(n+1)}}

\def\gvnow{\gv^{(n)}}
\def\gnow{g^{(n)}}

\def\gvplus{\gv^{(n+1)}}

\def\dphi{\delta\phi}
\def\dphinow{\delta\phinow}
\def\dphinowlm{\delta\phinowlm}

\def\Nxi{N_{\xi}}
\def\Nth{N_{\vartheta}}
\def\Nph{N_{\varphi}}

\def\hL {\hat{L}}
\def\hM {\hat{M}}
\def\dhMnow{\delta\hat{M}^{(n)}}
\def\dhMbar{\delta\hat{\mathcal{M}}^{(n)}}
\def\mubar{\bar{\mu}^{(n)}}
\def\munow{\mu^{(n)}}
\def\muprimenow{\mu^{\prime(n)}}
\def\muplus{\mu^{(n+1)}}

\def\pph {\phi}

\def\dphi{\delta\pph}

\slugcomment{Accepted, December 1, 2005}

\shorttitle{MOND density-potential pairs}
\shortauthors{Ciotti, Londrillo, \& Nipoti}

\begin{document}


\title{Axisymmetric and triaxial MOND density-potential pairs}


\author{L. Ciotti}
\affil{Dept. of Astronomy, University of Bologna, I-40127 Bologna, Italy}
\author{P. Londrillo}
\affil{INAF-Bologna Astronomical Observatory, I-40127 Bologna, Italy}
\author{C. Nipoti}
\affil{Dept. of Astronomy, University of Bologna, I-40127 Bologna, Italy}



\begin{abstract}

We present a simple method, based on the deformation of spherically
symmetric potentials, to construct explicit axisymmetric and triaxial
MOND density-potential pairs. General guidelines to the choice of
suitable deformations, so that the resulting density distribution is
nowhere negative, are presented. This flexible method offers for the
first time the possibility to study the MOND gravitational field for
sufficiently general and realistic density distributions without
resorting to sophisticated numerical codes.  The technique is
illustrated by constructing the MOND density-potential pair for a
triaxial galaxy model that, in the absence of deformation, reduces to
the Hernquist sphere.  Such analytical solutions are also relevant to
test and validate numerical codes. Here we present a new numerical
potential solver designed to solve the MOND field equation for
arbitrary density distributions: the code is tested with excellent
results against the analytic MOND triaxial Hernquist model and the
MOND razor-thin Kuzmin disk, and a simple application is finally
presented.

\end{abstract}


\keywords{gravitation --- stellar dynamics  --- galaxies: structure
  --- methods: analytical --- methods: numerical
   }


\section{Introduction}

Milgrom (1983) proposed that the failure of galactic rotation curves
to decline in Keplerian fashion outside the galaxies' luminous body
arises not because galaxies are embedded in massive dark halos, but
because Newton's law of gravity has to be modified for fields that
generate accelerations smaller than some characteristic value $\az$.
Subsequently, in order to solve basic problems presented by this
phenomenological formulation of the theory (now known as Modified
Newtonian Dynamics or MOND), such as conservation of linear momentum
(Felten 1984), Bekenstein \& Milgrom (1984) substituted the heuristic
1983 model with the MOND non-relativistic field equation
\begin{equation}
\nabla\cdot\left[\mu\left({\Vert\nabla\phi\Vert\over\az}\right)
                 \nabla\phi\right] = 4\pi G \rho,
\label{eqMOND}
\end{equation}
where $\Vert ...\Vert$ is the standard Euclidean norm, $\phi$ is the
gravitational potential produced by the density distribution $\rho$,
and $\nabla\phi\to 0$ for $\Vert\xv\Vert\to\infty$.  As stressed by
Bekenstein \& Milgrom (1984), the equation above is obtained from a
variational principle applied to a Lagrangian with all the required
symmetries, so the standard conservation laws are obeyed.  Thus,
equation~(\ref{eqMOND}) plays in MOND the same role as the Poisson
equation
\begin{equation}
\nabla^2\phiN=4\pi G\rho
\label{eqPoisson}
\end{equation}
in Newtonian gravity, and the MOND gravitational field $\gv$
experienced by a {\it test} particle is
\begin{equation} 
\gv=-\nabla\phi.
\label{eqgv}
\end{equation}

As well known, in the regime of intermediate accelerations the
function $\mu$ is not fully constrained by theory or observations,
while in the asymptotic regimes
\begin{equation}
\mu(t)\sim\cases{t&for $t\ll 1$,\cr 1&for $t\gg 1$. }
\label{eqmuasym}
\end{equation}
Throughout the paper we conform to the standard
assumption
\begin{equation}
\mu (t)={t\over\sqrt{1+t^2}}
\label{eqmu}
\end{equation}
(see however Famaey \& Binney 2005).

From equation~(\ref{eqmuasym}) it follows that equation~(\ref{eqMOND})
reduces to the Poisson equation when $\Vert \nabla \phi \Vert \gg
\az$, while the limit equation 
\begin{equation}
\nabla\cdot\left({\Vert\nabla\phi\Vert}\nabla\phi\right) = 4\pi G \az
\rho,
\label{eqdMOND}
\end{equation}
obtained by assuming $\mu(t)=t$ in equation~(\ref{eqMOND}), describes
systems for which (or regions of space where) $\Vert\nabla\phi\Vert
\ll\az$, i.e. systems for which the MOND predictions differ most from
the Newtonian ones.  As a consequence equation~(\ref{eqdMOND}),
characterizing the so-called `deep MOND regime' (hereafter dMOND), is
of particular relevance in MOND investigations.

Nowadays a considerable body of observational data seems to support
MOND well beyond its originally intended field of application (see,
e.g., Milgrom~2002; Sanders \& McGaugh~2002), making this theory an
interesting alternative to the Cold Dark Matter paradigm. It is thus
natural to study in detail MOND predictions, in particular focusing on
dMOND systems, i.e. systems that should be dark matter dominated if
Newtonian gravity holds.  Potential problems of the theory have
already been pointed out by various authors (see, e.g. The \& White
1988; Buote et al.~2002; Sanders~2003; Ciotti \& Binney~2004; Knebe \&
Gibson~2004; Zhao et al.~2005), but further analyses are needed to
reach firmer conclusions.  Unfortunately, MOND investigations have
been considerably slowed down by the lack of aspherical
density-potential pairs to test theory predictions in cases more
realistic than those described by spherical symmetry: the search for a
sufficiently general method to construct aspherical MOND solutions is
the subject of this paper.

In MOND, the main difficulty to obtain exact aspherical
density-potential pairs (or to build robust numerical solvers)
originates from the non-linear nature of the theory, which makes
impossible a straightforward use of the analytical and numerical
techniques available for the Poisson equation (such as integral
transforms or expansion in orthogonal functions).  In addition, a
simple relation between the Newtonian and the MOND gravity fields in
general does not exist. Although equation~(\ref{eqPoisson}) can be
used to lower the order of equation~(\ref{eqMOND}) by eliminating the
source density, it follows that
$\mu(\Vert\nabla\phi\Vert/\az)\nabla\phi$ and $\nabla\phiN$ differ by
some unknown solenoidal field\footnote{That the solenoidal field $\Sv$
in general cannot be arbitrarily set to zero is due to the fact that
the MOND acceleration field must be derived from a potential, while
equation~(\ref{eqmug}), if correct in general, would imply that
$\nabla \Vert \gv \Vert \wedge \gv=0$, an identity which is not
necessarily true when $\gv$ is derived from a potential.} $\Sv={\rm
curl\,}\hv$. Remarkably, Brada \& Milgrom (1995; hereafter BM95)
showed that when the modulus $\gN$ of the Newtonian gravitational
field produced by $\rho$ is a function of $\phiN$ only, $\Sv$
vanishes, and so the MOND acceleration $\gv$ is related to
$\gvN\equiv-\nabla\phiN$ by
\begin{equation}
\mu\left({g\over\az}\right)\gv=\gvN,
\label{eqmug}
\end{equation}
where $g=\Vert\gv\Vert$.  Equation~(\ref{eqmug}) coincides with the
original MOND formulation of Milgrom (1983) and can be solved
algebraically: particularly simple cases described by this equations
are those in which the density distribution is spherically or
cylindrically symmetric, or stratified on homogeneous planes.  In such
cases the MOND potential is not more difficult to construct than the
corresponding Newtonian potential.  This is the reason why {\it all}
the (astrophysically relevant) analytical MOND density-potential pairs
known are spherically symmetric: in fact, the only exact, aspherical
MOND density-potential pair available is the axisymmetric razor-thin
Kuzmin disk (and the derived family; BM95), for which
$\gN=\gN(\phiN)$.  In all the other cases one is forced to solve
numerically equation~(\ref{eqMOND}).

The importance of a general method to obtain the explicit MOND
potential of density distributions with prescribed shape and
stratification is then obvious, because it would allow for orbit
integration, without resorting to numerical integration of
equation~(\ref{eqMOND}); in addition, analytical solutions could be
used to test numerical MOND solvers in more realistic cases than
spherical symmetry.  In this paper we show that such an approach can
be devised. In particular, we show how a ``seed'' spherical density
distribution can be deformed in an axisymmetric or triaxial density
distribution with analytical potential satisfying the MOND equation,
by means of a pair of very simple existence theorems and (for example)
by using building blocks obtained from the Ciotti \& Bertin (2005;
hereafter CB05) method. In addition, as a complement to the analytical
method, we also illustrate a new numerical MOND solver based on
spectral methods, which adds to the short list of the others available
(Milgrom 1986; BM95; see also Brada \& Milgrom~1999).

The paper is organized as follows. In Section~2 we present the general
method (postponing to the Appendix the proof of three technical
results on which the method is based), and in Section~3 we construct
families of triaxial profiles obtained by deformation of the Hernquist
(1990) sphere. The original numerical code developed to compute the
MOND potential of generic density distributions is then described in
Section~4, where we also show how well it recovers the analytical
triaxial potentials of Section~3 and the MOND Kuzmin density-potential
pair. The code is finally used to investigate the field $\Sv$ of
highly flattened triaxial density distributions.  The main results of
the paper and possible future applications are finally summarized in
Section~5.

\section{The general method}
\label{genmeth}

Two different approaches are possible to construct explicit solutions
of equation (\ref{eqMOND}): in the most obvious one attempts to
recover the potential of a given density $\rho$.  Unfortunately, no
explicit aspherical density-potential pairs (with the exception of the
Kuzmin disk) have been obtained so far following this approach. In
addition, we note that even ``innocent'' density distributions can
produce puzzling behaviors of the MOND acceleration field at special
places, as illustrated by the perturbative case analyzed by Ciotti \&
Binney~(2004).

In the alternative ``$\phi$-to-$\rho$'' approach -- studied in this
paper -- one determines the density field by application of the MOND
operator to a prescribed potential\footnote{A first attempt to
construct the MOND potential of an oblate galaxy along this line was
carried out by Hongsheng Zhao (private communication).}.  This
approach is straightforward because only differentiation is involved,
but negative densities can be produced if the potential is chosen
without the necessary care, as well known also in the Newtonian
case. Thus, a first problem posed by this approach is to find a
criterion to choose $\phi$ such that $\rho$ is positive.  In addition
to the positivity condition, one would like to be able to control the
{\it shape} and the {\it radial trend} of the resulting density
distribution: for example, in Newtonian gravity homeoidally stratified
potentials lead to curious toroidal density distributions
(e.g. Binney~1981, Evans~1994). As we will see, it is possible -- to
some degree -- to answer positively to these issues.

The idea behind our method is to compute the MOND potential of a
spherical density distribution. The obtained potential is then mapped
on a new spherical density by means of the Laplace operator. This new
density is then suitably deformed and the associated (Newtonian)
potential is determined by standard methods.  Finally, the MOND
differential operator is applied to the deformed potential, and the
final density distribution is obtained. In practice, the method
consists of the following steps.

\begin{figure}
\epsscale{.70}
\plotone{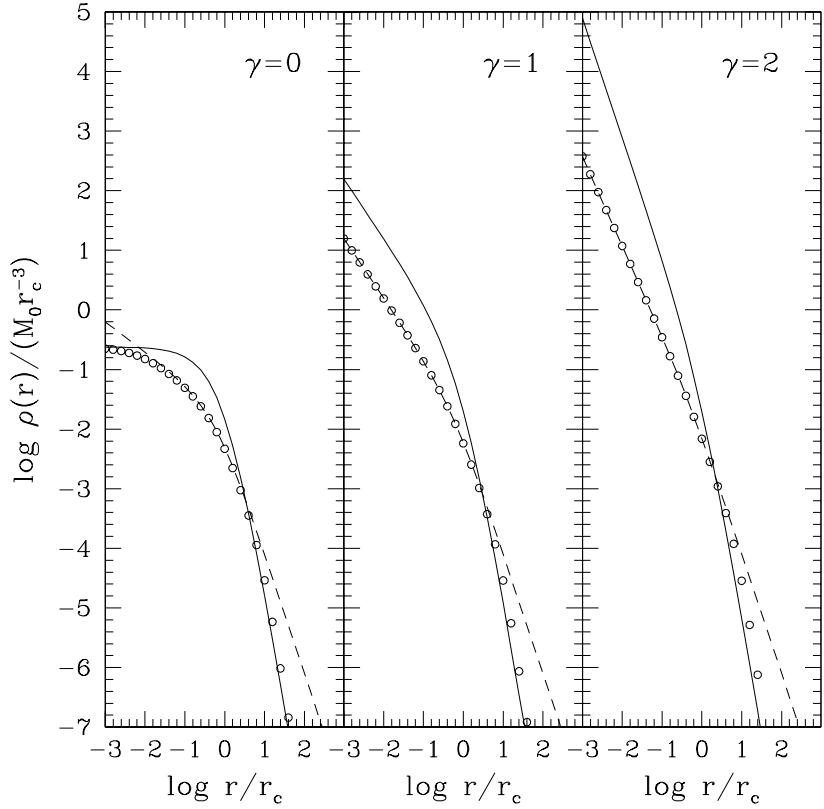}
\caption{The Newtonian (dashed lines) density profile $\rhoNz$
[equation~(\ref{eqrhoNz})] and the full MOND (empty symbols) density
profile $\rhoMz$ [equation~(\ref{eqrhoMz})] derived from the dMOND
potential of the spherical $\gamma$-model
$\rhoz=(3-\gamma)\Mz\rc/[4\pi r^{\gamma} (r+\rc)^{4-\gamma}]$ for
$\gamma=0,1,2$ (solid lines).  In all cases we adopted $G
\Mz=100\az \rc^2$. }
\label{figgamma}
\end{figure}

1) We start by selecting a spherical density distribution $\rhoz(r)$
   with the desired radial behavior: for example, it could be a
   $\gamma$-model (Dehnen~1993; Tremaine et al.~1994), a King~(1972)
   density distribution, or a classical polytrope such as the
   Plummer~(1911) sphere.  We then calculate its dMOND potential
   $\phiz(r)$ by using equation~(\ref{eqmug}) in spherical symmetry:
\begin{equation}
{d\phiz\over dr}=\gz={\sqrt{G\Mz(r)\az}\over r},
\label{eqgz}
\end{equation}
   where $\Mz(r)=4\pi\int_0^r\rhoz(r)r^2dr$. As is well known,
   $\phiz(r\to\infty)\sim\sqrt{G\Mz\az}\ln r$ for any mass
   distribution of finite total mass $\Mz$.  We stress that we are
   calculating the dMOND potential of $\rhoz$ irrespective of whether
   the density distribution is actually dMOND everywhere: for example,
   the Jaffe~(1983) model near the center is not dMOND because its
   acceleration field diverges there, yet we can integrate
   equation~(\ref{eqgz}) to obtain $\phiz$. The reasons for doing so
   are that equation~(\ref{eqgz}) is much easier to integrate than the
   corresponding MOND (spherically symmetric) equation, and that
   Theorem~2 assures that both the Laplacian and the MOND operator
   applied to $\phiz$ will produce positive densities. In particular,
   the Laplace operator applied to $\phiz$ leads to the density
   distribution
\begin{equation}
\rhoNz=\sqrt{\az\over G}\left[{\rhoz(r)r \over 2\sqrt{\Mz(r)}}
+{\sqrt{\Mz(r)} \over 4 \pi r^2} \right],
\label{eqrhoNz}
\end{equation}
  while the MOND operator in equation~(\ref{eqMOND}) produces 
\begin{equation}
\rhoMz=\mu\left({\gz\over\az}\right)\left[\rhoNz+{\gz^{\prime}\over 4\pi G \left(1+\gz^2/\az^2\right)}\right].
\label{eqrhoMz}
\end{equation}
  By construction $\phiz$ is the dMOND potential of $\rhoz$, the
  Newtonian potential of $\rhoNz$, and the MOND potential of $\rhoMz$.
  Obviously, $\rhoMz$ coincides with $\rhoz$ where the model is in
  dMOND regime, and with $\rhoNz$ where the gravity field $\gz$ is
  strong enough. From a more quantitative point of view,
  $\rhoNz(r\to\infty)\propto r^{-2}$ when the total mass of $\rhoz$ is
  finite, and $\rhoNz(r\to0)\propto r^{-(1+a)/2}$ if $\rhoz \propto
  r^{-a}$ $(a < 3)$ in the central regions.  This is apparent from
  Fig.~\ref{figgamma}, where we plot $\rhoz$ (solid lines), $\rhoNz$
  (dashed lines) and $\rhoMz$ (empty symbols) for $\gamma$-models with
  $\gamma=0$, 1, and 2.

2) In the second step of the procedure we look for an aspherical
   density $\rhoNu$ to be added to $\rhoNz$ so that
   $\rhoNz+\lambda\rhoNu$ is positive for some $\lambda>0$, and
   $\phiu$ (defined by $\nabla^2\phiu=4 \pi G \rhoNu$) is explicitly
   known. The associated total (Newtonian) potential is thus
   $\phi=\phiz+\lambda\phiu$. The main reason for working with
   Newtonian potential at this stage is that the properties of
   density-potential pairs obeying the Poisson equation are well
   known, and so we can work with confidence and control the
   deformations of the total $\rhoNz+\lambda\rhoNu$.
  
3) The final step is the explicit evaluation of the dMOND operator in
   equation~(\ref{eqdMOND}) for the potential
   $\phi=\phiz+\lambda\phiu$, and this can be done with computer
   algebra packages such as Maple or Mathematica.  For $\lambda=0$ one
   recovers $\phiz$ (which produces the positive $\rhoz$), and for
   continuity one could expect that a positive $\rho$ (with a radial
   trend similar to that of $\rhoz$) is also obtained for sufficiently
   small $\lambda$, though we will show that this is not always the
   case\footnote{See the discussion below equation~(\ref{eqpsid}).}.
   In general, for given $\rho$ and $\rhoNu$ a critical value of
   $\lambda$ exists such that only smaller values $\lambda$ correspond
   to physically acceptable (i.e. nowhere negative) $\rho$.  If the
   resulting density is positive, then Theorem~1 assures us that also
   the full MOND operator applied to $\phi$ will produce a positive
   density distribution (such that its limit for $\lambda\to0$ is
   $\rhoMz$), provided that $\rhoNz+\lambda\rhoNu \geq 0$. Note that
   the last requirement is just a {\it sufficient} condition for the
   positivity of $\rho$.
   
   A first hint about the positivity and the shape of the resulting
   density can be obtained by studying the linearization of the dMOND
   operator. In fact, it is easy to prove that the
   $\lambda$-linearized density associated with
   $\phi=\phiz+\lambda\phiu$ is
\begin{equation}
\rho=\rhoz +{\lambda\over 4\pi G\az}\left[
            \left(4\pi G\rhoNu +{\partial^2\phiu\over\partial r^2}\right)\gz+
            \left(4\pi G\rhoNz +{d^2\phiz\over
            dr^2}\right){\partial\phiu\over\partial r} \right]+O(\lambda^2),
\label{eqlinrho}
\end{equation} 
where the spherical symmetry of $\phiz$ has been exploited by using
the standard spherical coordinates ($r$, $\vartheta$, $\varphi$), and
it is intended that $\phiu= \phiu(r,\vartheta,\varphi)$ and $\gz\neq 0$.

From the discussion above it should be obvious that the delicate step
in the procedure is the choice of $\rhoNu$ such that
$\rhoNz+\lambda\rhoNu$ is nowhere negative, $\phiu$ is analytical, and
the dMOND operator produces the sought deformation on $\rho$.  Simple
choices satisfying the first two requests could be the addition of
Miyamoto \& Nagai (1975) or Satho (1980) disks: however, an even more
general approach can be devised.  In fact, families of explicit and
easy-to-calculate Newtonian $(\rhoNu ,\phiu)$ pairs can be obtained
from the homeoidal expansion technique (e.g., see CB05).  In practice,
for the present problem one can adopt as $\rhoNu$ the linear term in
the homeoidal expansion of a {\it seed density} $\varrho$, i.e.
\begin{equation}
\rhoNu=(\epsilon y^2 +\eta z^2){\varrho^{\prime}(r)\over r}, 
\label{eqrhoNu}
\end{equation} 
where the dimensionless parameters $\epsilon$ and $\eta$ are the
flattenings of the expanded homeoidal density distribution.  The
potential $\phiu$ generated by the distribution $\rhoNu$ is written
in terms of one-dimensional radial integrals that usually can be
evaluated explicitly (see equations [5] and [6] in CB05):
\begin{equation}
{\phiu \over 4\pi G}=(\epsilon+\eta)\psiu(r)+(\epsilon y^2 +\eta z^2)\psid(r),
\label{eqphiu}
\end{equation} 
where
\begin{equation}
\psiu(r)=\int_0^r{{\varrho(m)m^2 \over r}\left(1-{m^2
  \over 3 r^2 }\right) \de m}+{2 \over 3}\int_r^\infty{\varrho(m)m
  \de m},
\label{eqpsiu}
\end{equation} 
and
\begin{equation}
\psid(r)={1 \over r^5}\int_0^r{\varrho(m)m^4\de m}.
\label{eqpsid}
\end{equation}

We stress that in the present application $\epsilon$ and $\eta$ are
just linear parameters, so without loss of generality the
multiplicative coefficient $\lambda$ can be considered to be contained
in them, and their value is not restricted by the limitations
described in appendix A of CB05. Note also that $\rhoNu$ in
equation~(\ref{eqrhoNu}) is {\it negative} when it is derived from an
{\it oblate} axisymmetric (the $\epsilon=0$ case) or triaxial
$\varrho$, and so particular care is needed in the choice of
$\varrho$, in order to have $\rhoNz+\rhoNu\geq 0$. However, for a
$\varrho$ that can be approximated by a power-law in its external
regions, $\rhoNu(r\to \infty)\sim -\varrho$, and so any finite-mass
$\varrho$ produces through equation~(\ref{eqrhoNu}) a positive total
density at large radii, where $\rhoNz \sim r^{-2}$.  Finally we remark
that, for $\rhoNu$ as in equation~(\ref{eqrhoNu}), the quantities
${\partial \phiu / \partial r}$ and ${\partial^2 \phiu / \partial
r^2}$ in equation~(\ref{eqlinrho}) are particularly simple because,
according to equation~(\ref{eqphiu}), the angular part of $\phiu$ is
just a multiplicative factor of the radial function $r^2\psid$.

We note that the seemingly obvious choice of taking $\rhoNu$ to be the
homeoidal expansion term derived from $\rhoNz$, does not work. This is
because $\rhoNz \sim r^{-2}$ at large radii, so $\rhoNu \propto
-(\epsilon\sthsq\sphsq+\eta\cthsq)/r^2$ and $\phiu \propto
-2(\epsilon+\eta)\ln r /3+ (\epsilon\sthsq\sphsq+\eta\cthsq)/3$ for
$r\to \infty$: equation~(\ref{eqlinrho}) then reveals that the density
is negative for sufficiently large $r$ in the conical region $\cth <
1/\sqrt{3}$.  As we will see, the regions near the $z$ axis are in
fact the most delicate when applying the analytical technique
presented in this paper: unphysical models usually develop negative
densities near the $z$-axis.

We conclude this Section by noting that, when truncated at the first
order in $\lambda$, $\phiz+\lambda\phiu$ and the $\lambda$-linearized
$\rho$ in equation~(\ref{eqlinrho}) allow for the explicit evaluation
of the $\lambda$-linearized solenoidal field $\Sv=\Vert \nabla \phi
\Vert \nabla \phi / \az - \nabla\phiN=\lambda \Svu +O(\lambda^2)$ (the
zeroth-order term in $\lambda$ is obviously missing because it refers
to the spherical density component). If $\phiu$ is given by
equation~(\ref{eqphiu}) then $\rho$ in equation~(\ref{eqlinrho})
belongs to the family considered in CB05 and its Newtonian potential
can be derived in closed form (while in general this is not possible
when considering the density obtained by the application of the dMOND
operator without $\lambda$-linearization).

\begin{figure}
\epsscale{1.1}
\centerline{
\plottwo{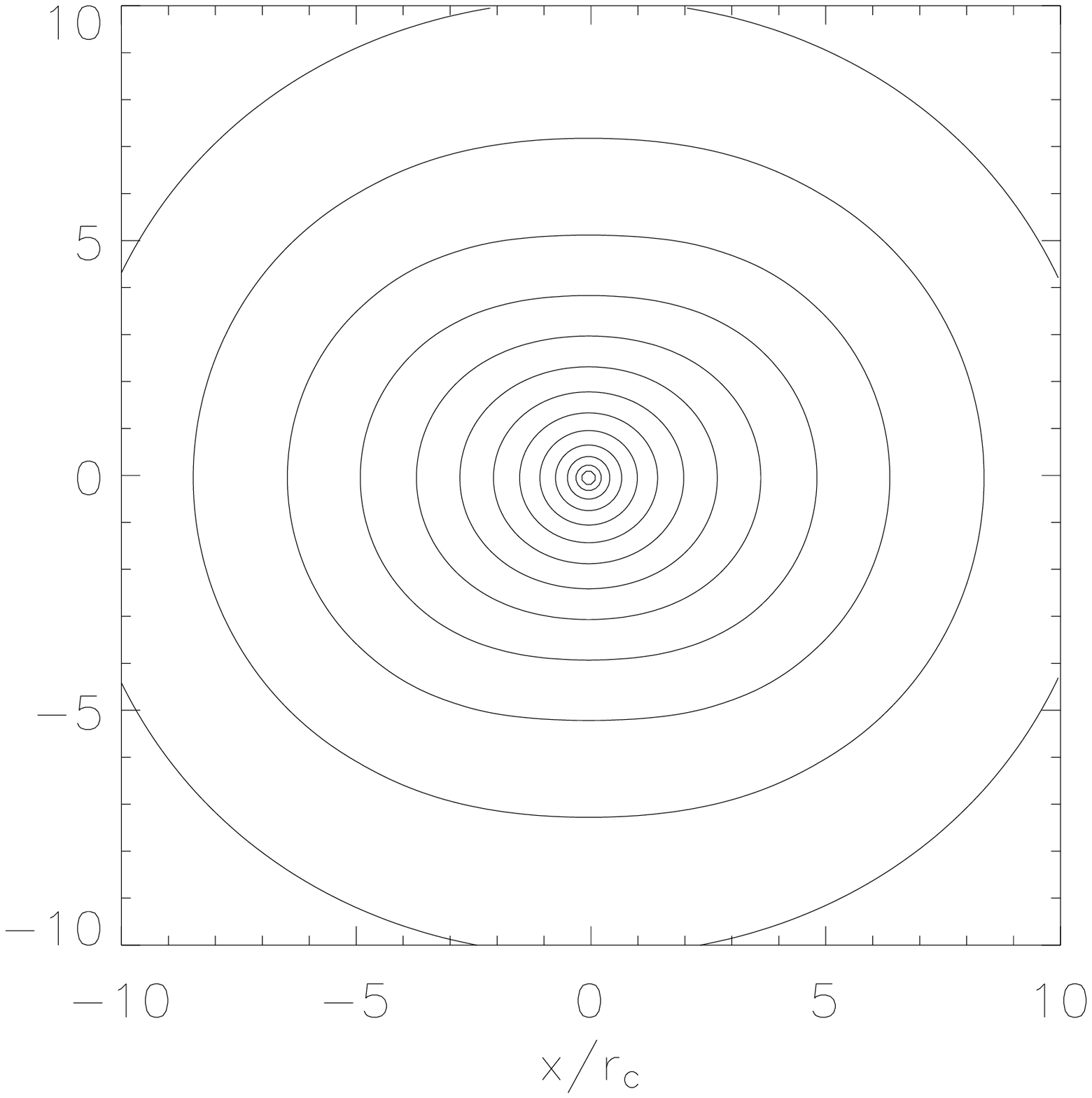}{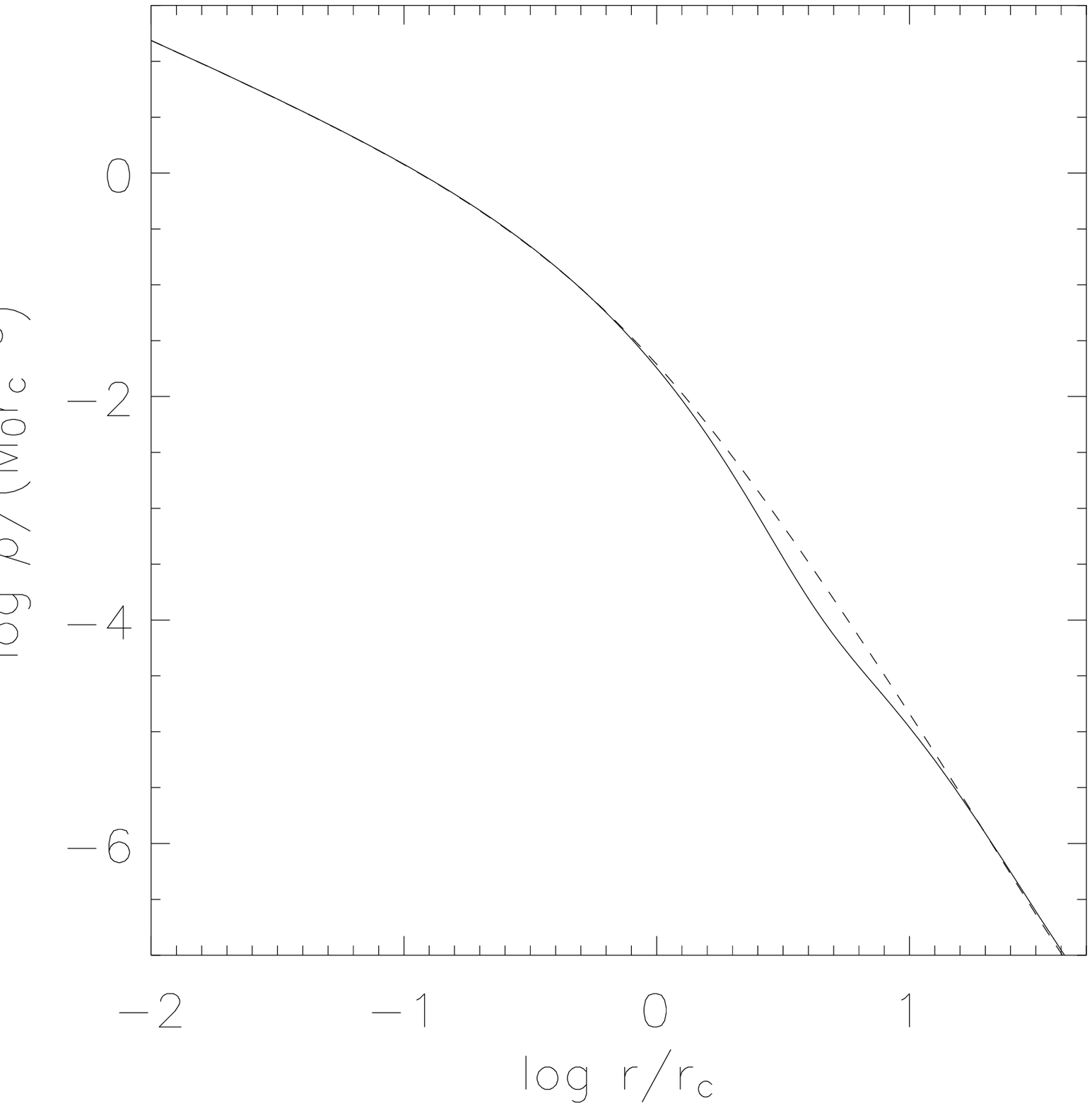}}
\centerline{
\plottwo{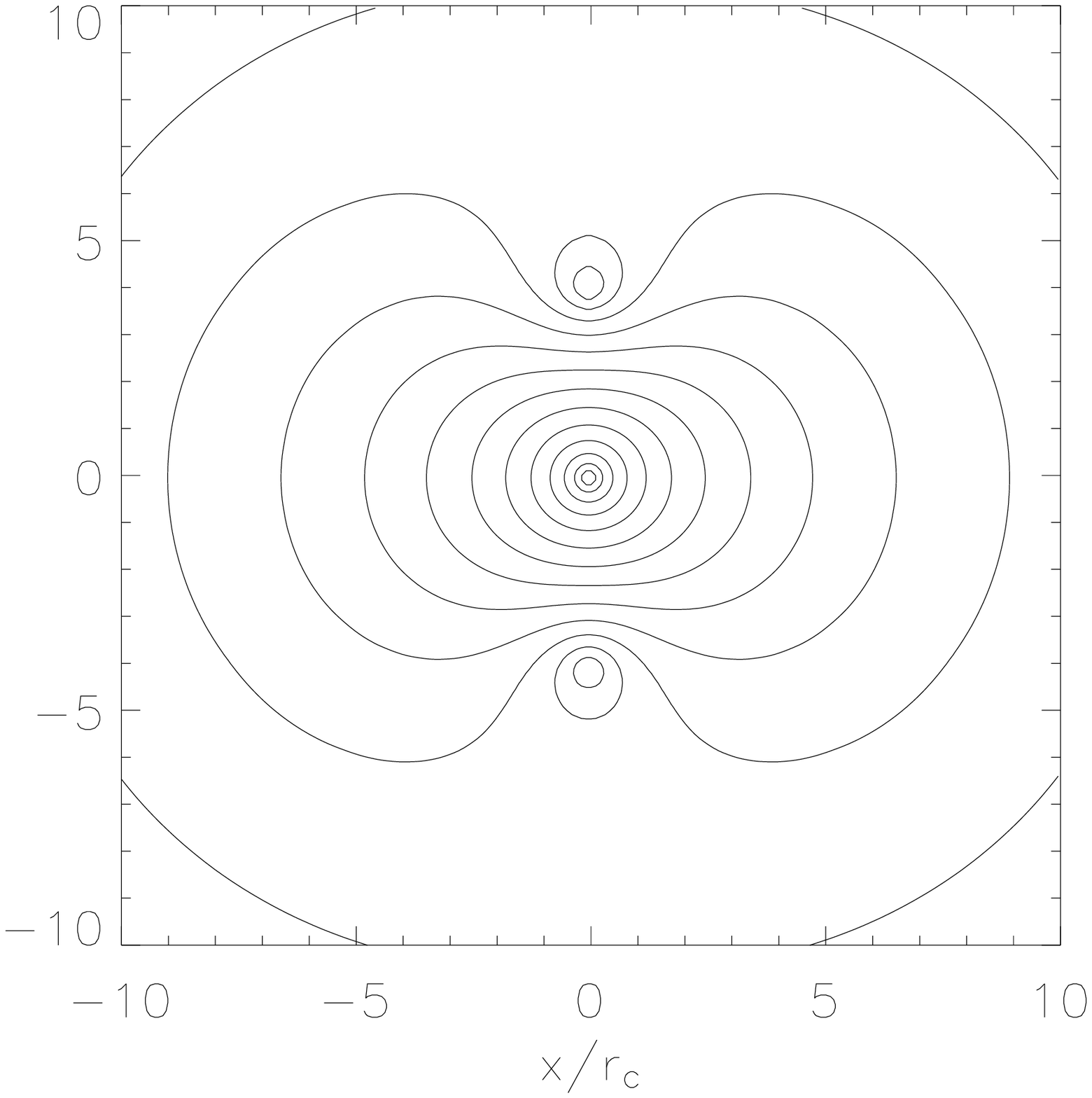}{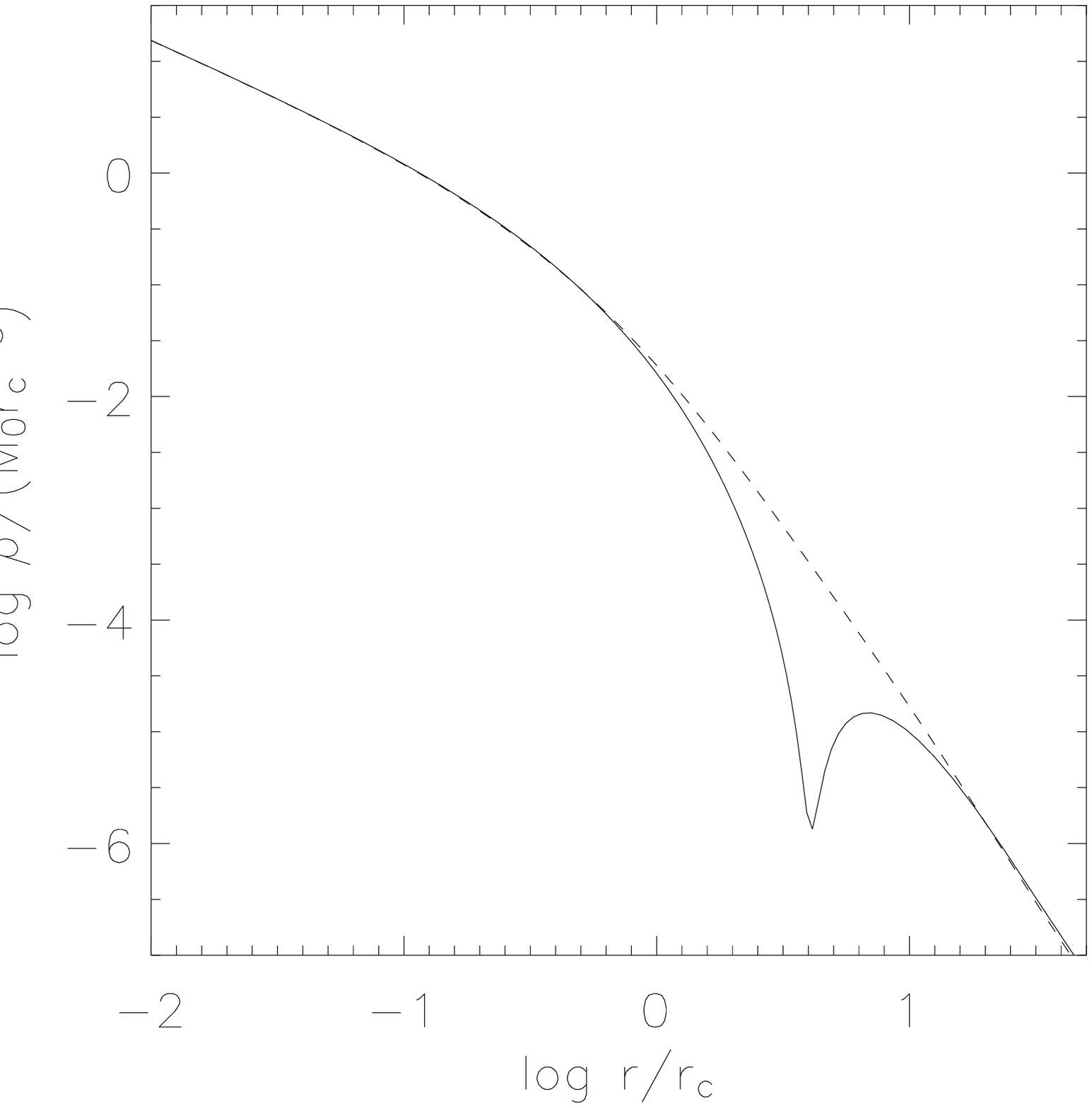}}
\caption{Isodensity contours (left panels) and density profiles (right
  panels) for two analytical dMOND axisymmetric ($\epsilon=0$)
  Hernquist models with $\phiu$ as in equation~(\ref{eqphiusix}). The
  density profiles are taken along a radius in the equatorial plane
  (dashed lines) and along the symmetry axis $z$ (solid lines). The
  model in the top panels has $\beta=5$ and $\etatilde=0.01$, while
  the model in the bottom panels has $\beta=5$ and $\etatilde=0.02$.}
\label{figden}
\end{figure}

\begin{figure}
\epsscale{1.1}
\centerline{
\plottwo{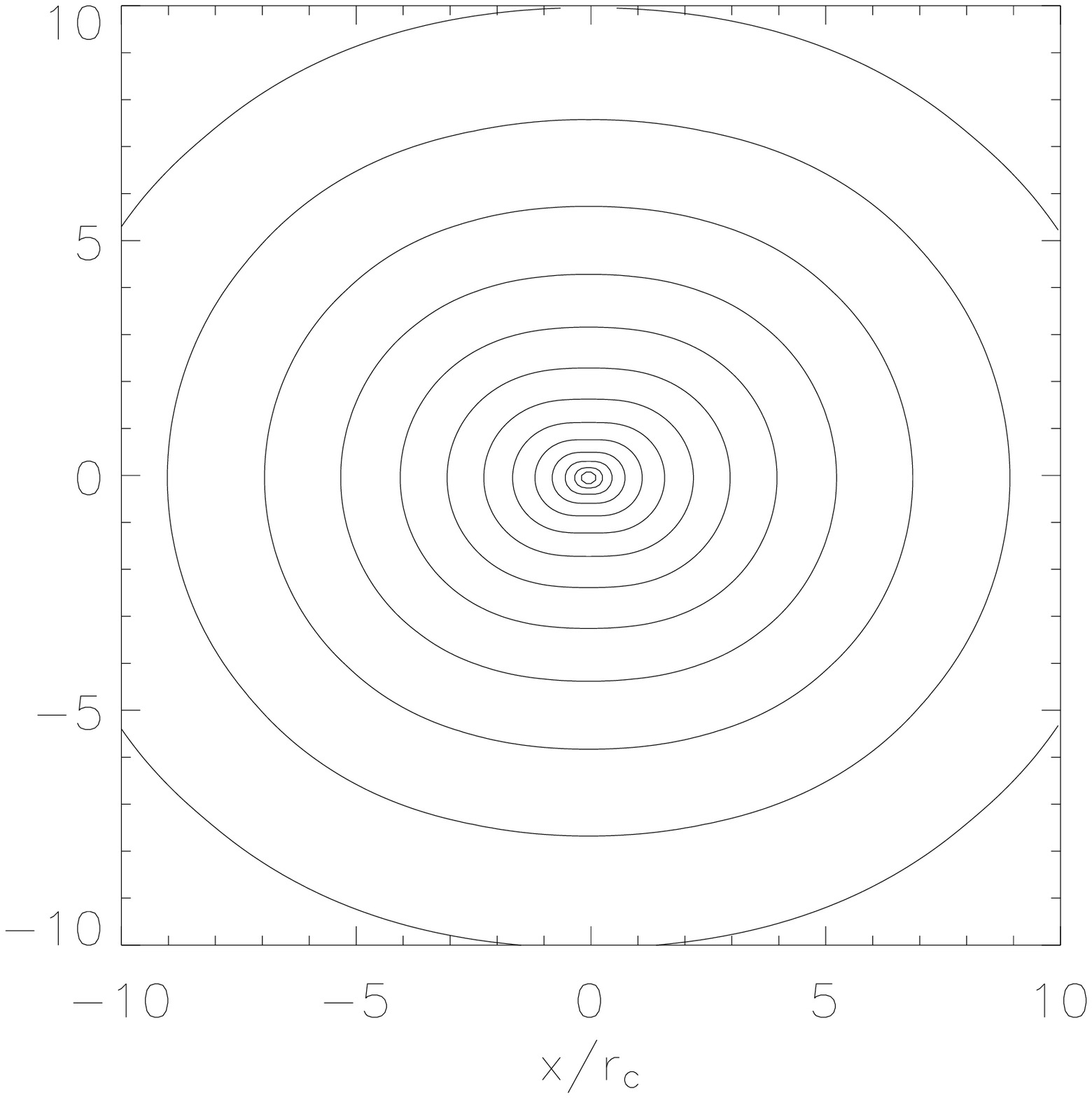}{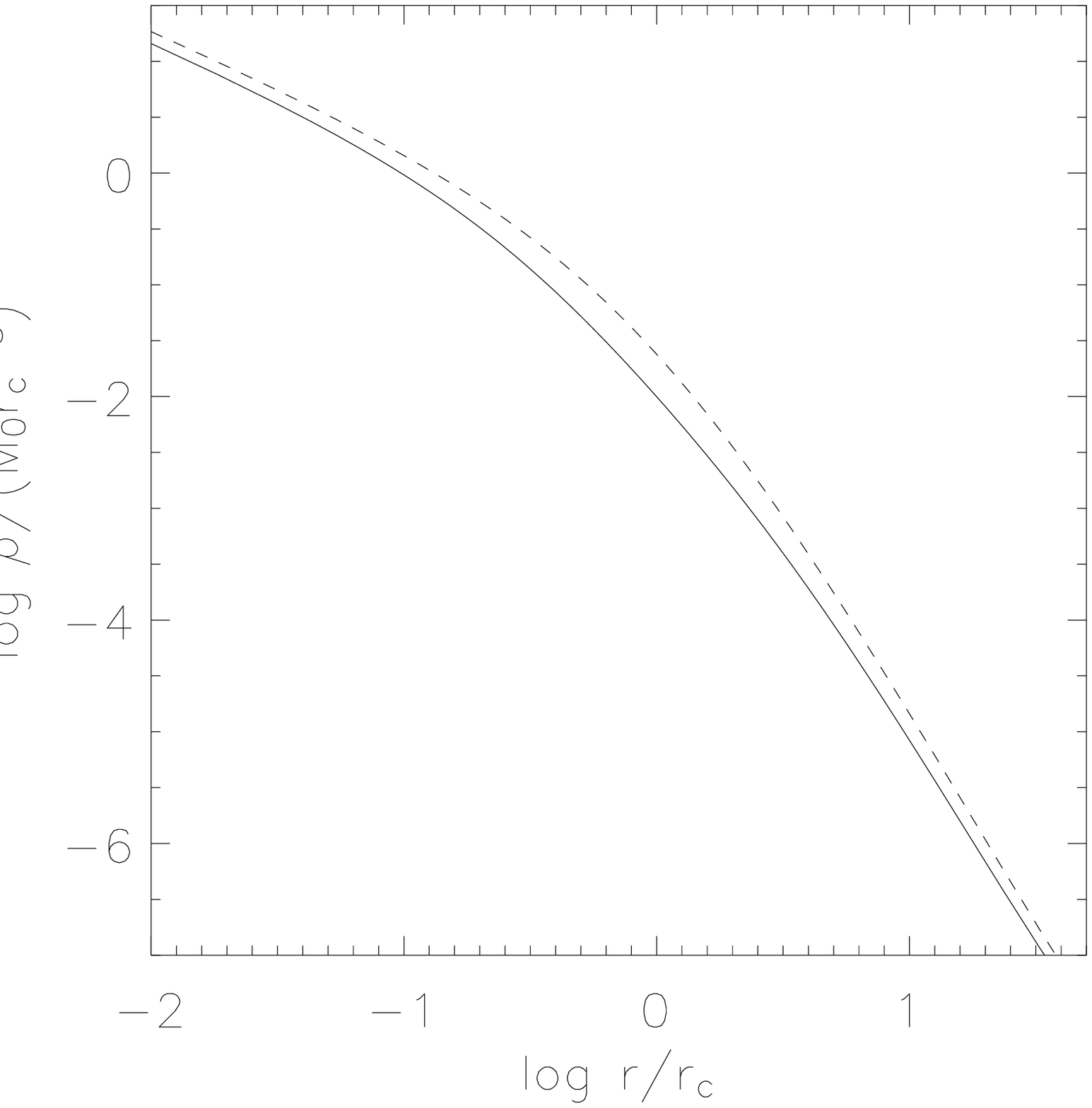}}
\centerline{
\plottwo{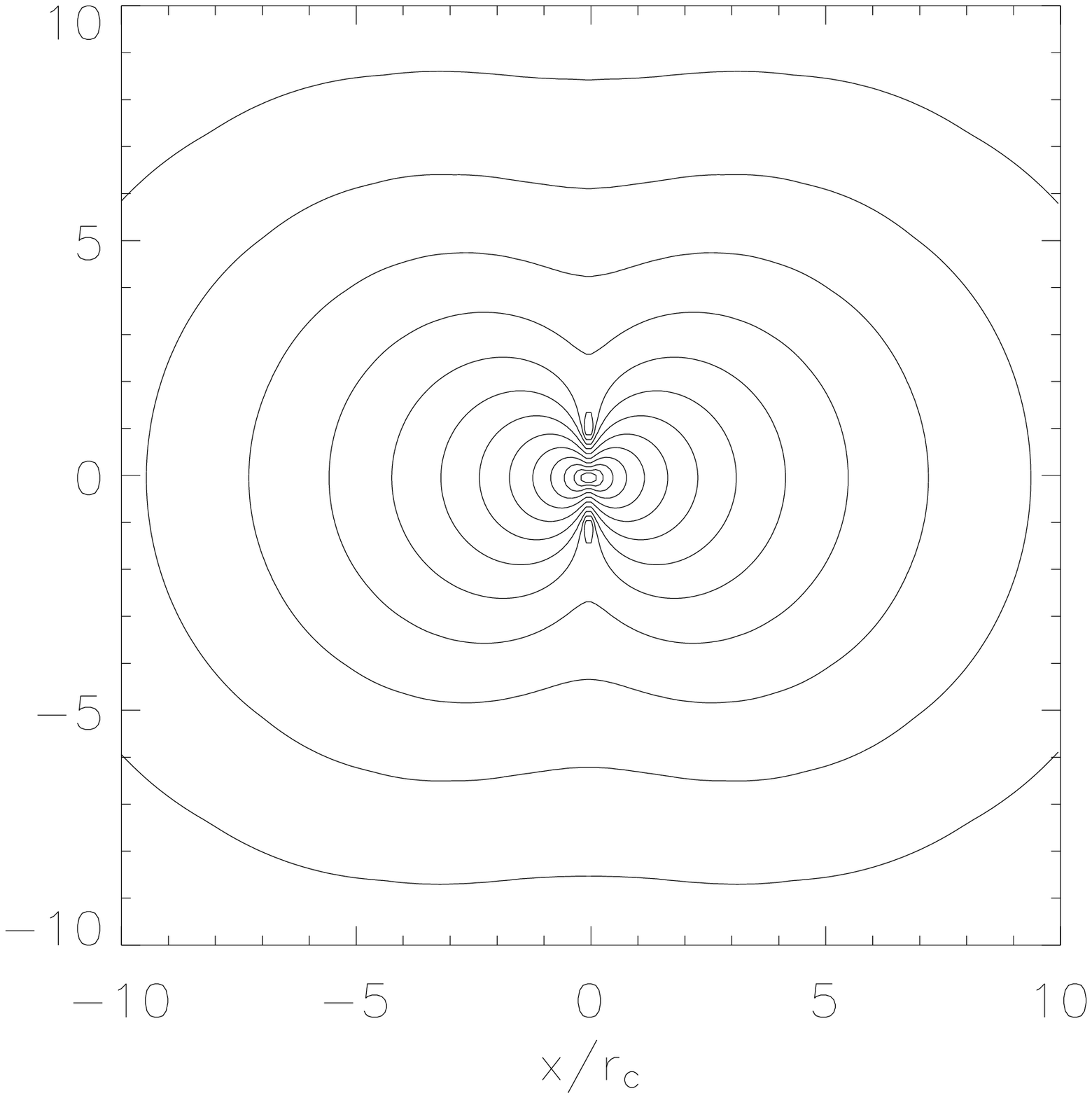}{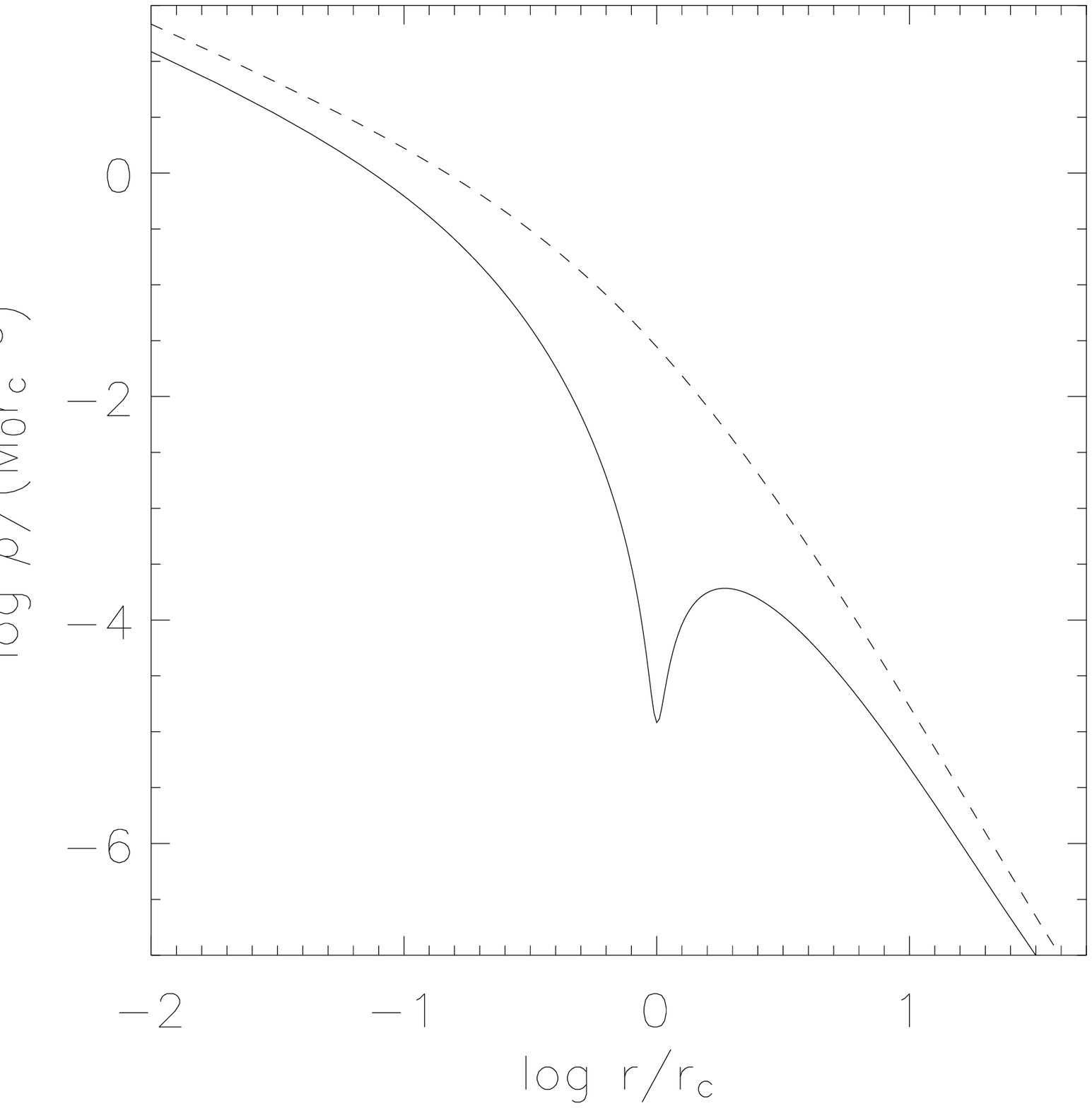}}
\caption{The same quantities as in Fig.~\ref{figden} for two
  analytical dMOND axisymmetric ($\epsilon=0$) Hernquist models
  with $\phiu$ as in equation~(\ref{eqphiuq}), with $\eta=0.2$ (top)
  and $\eta=0.4$ (bottom).}
\label{figdenbis}
\end{figure}

\begin{figure}
\epsscale{.9}
\centerline{
\plottwo{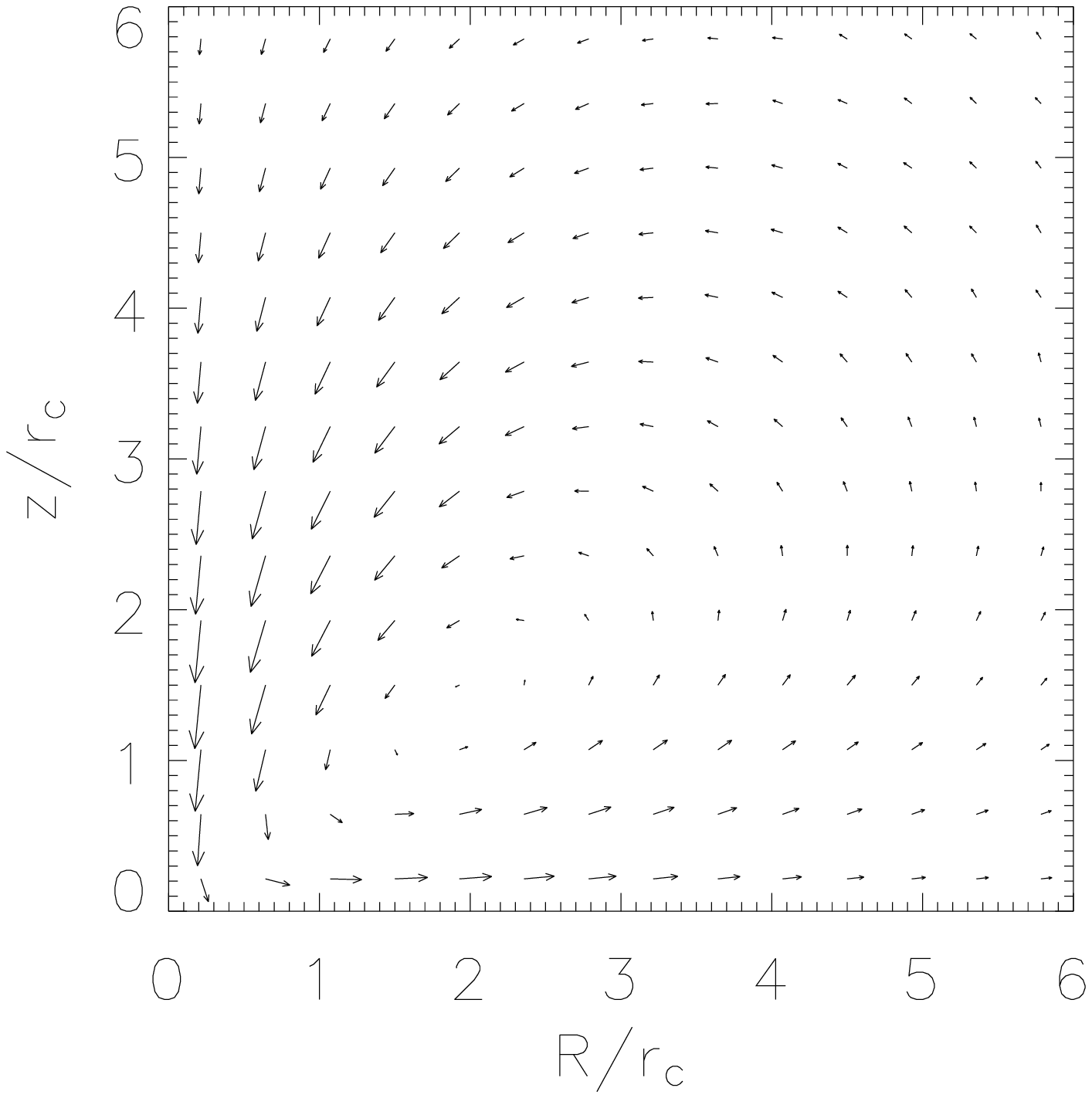}{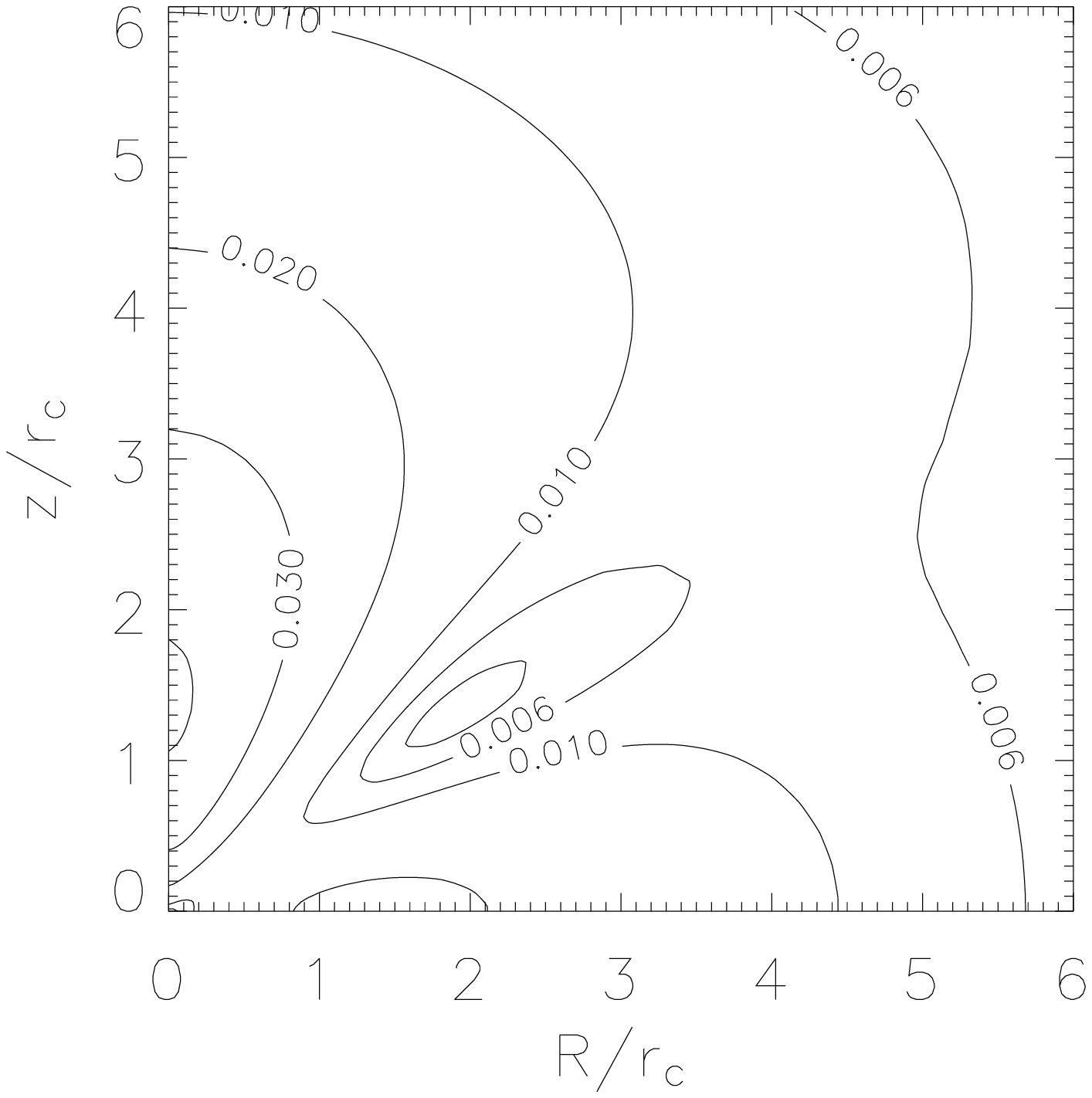}}  %
\centerline{
\plottwo{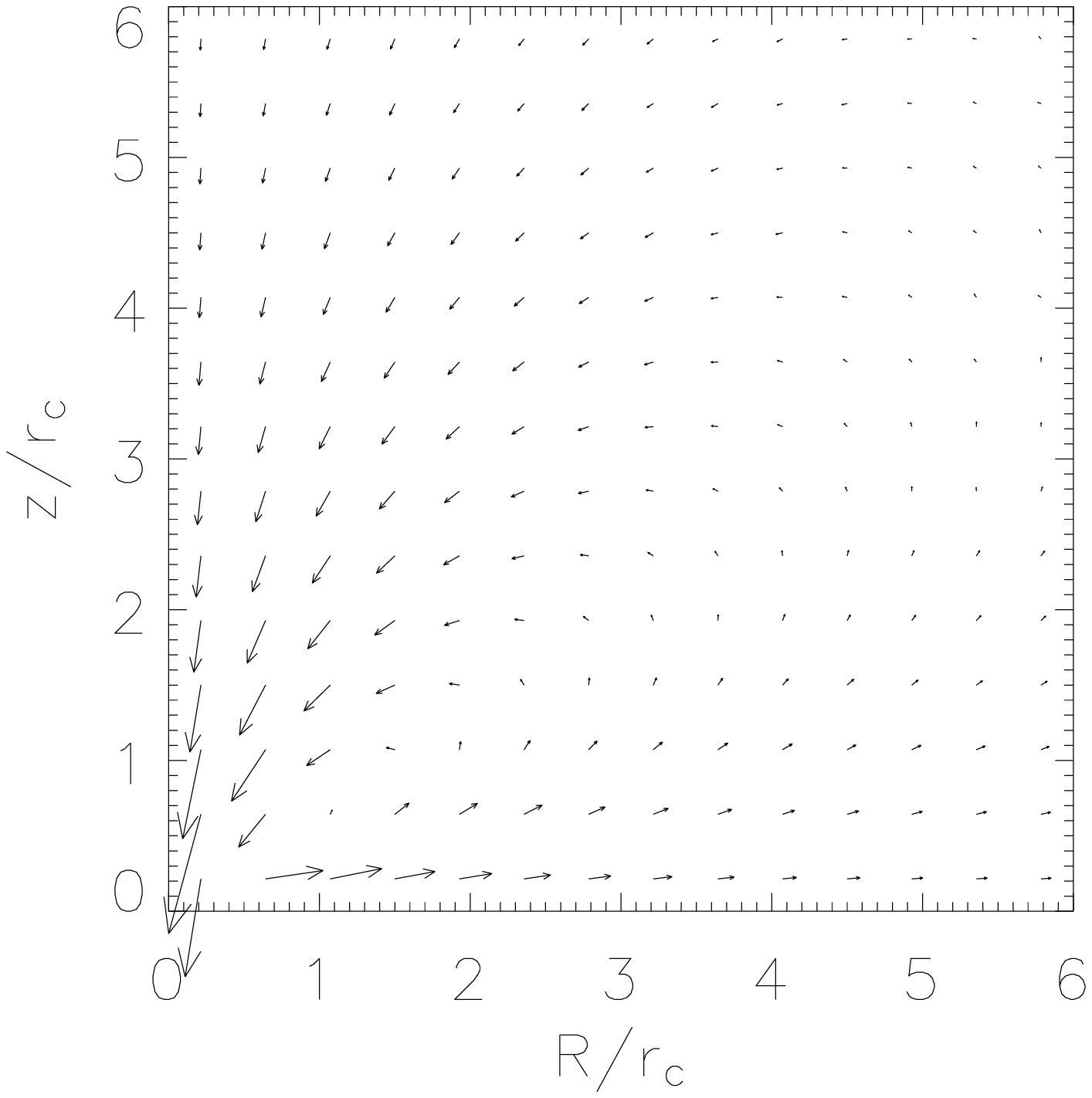}{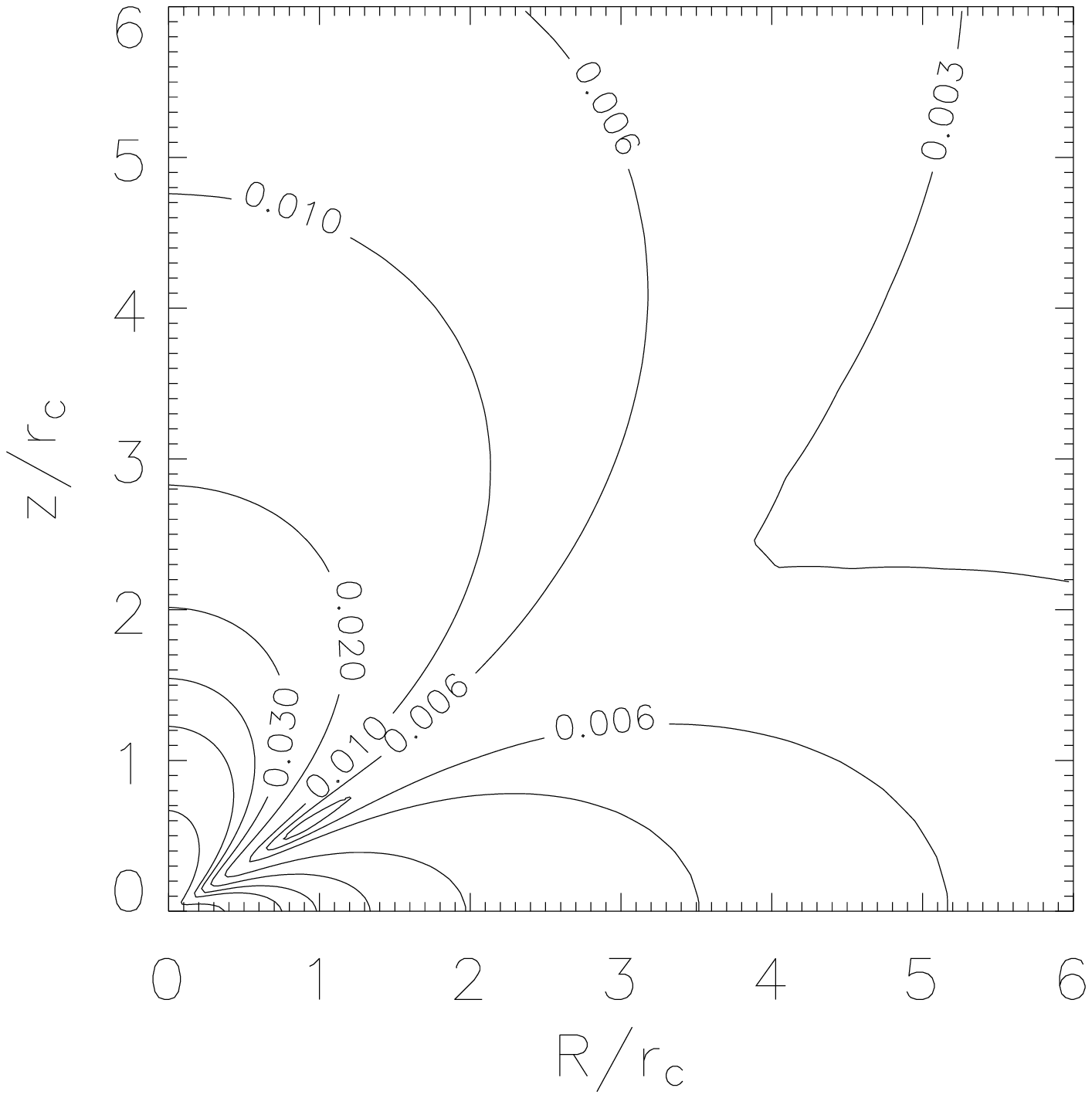}
}
\centerline{
\plottwo{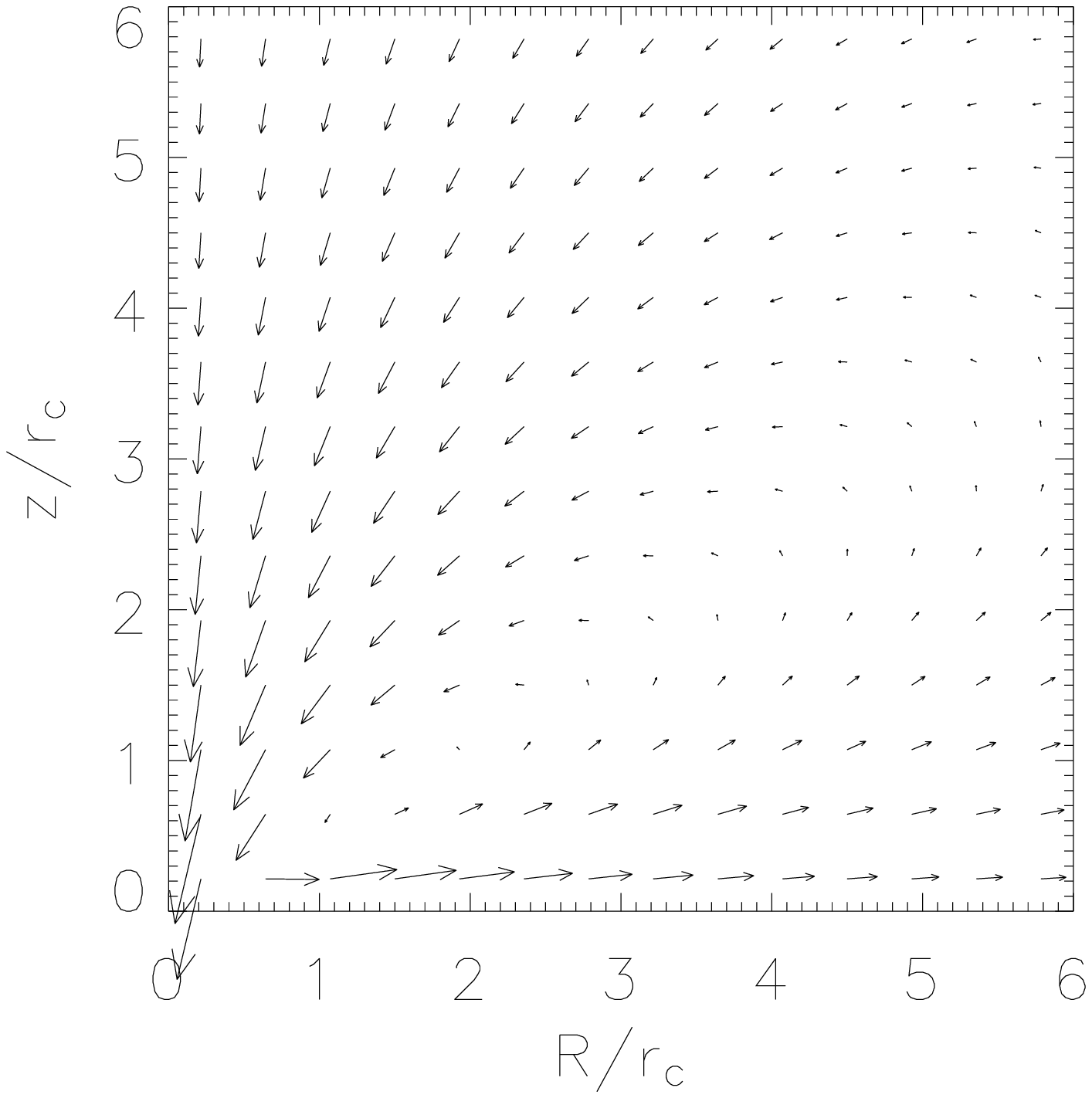}{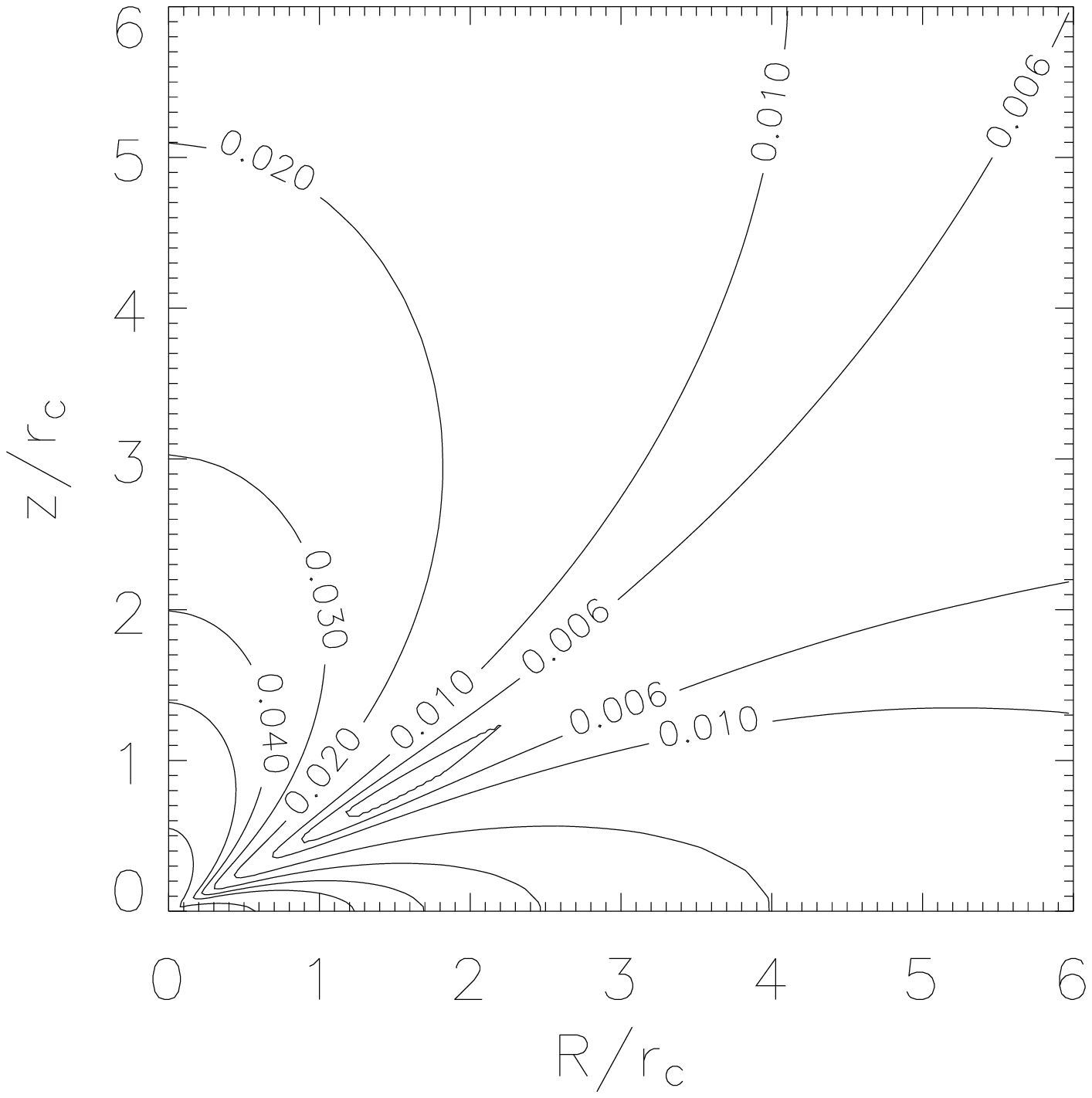}
}
\caption{The dMOND solenoidal field $\Sv\equiv g \gv/\az-\gvN$
(normalized to $\gN$) in the meridional plane for three different
axisymmetric Hernquist models (left panels). In the right panels lines
of constant $\Vert \Sv \Vert/\gN$ are shown. From top to bottom: the
model of Fig.~\ref{figden} with $\etatilde=0.02$, the model of
Fig.~\ref{figdenbis} with $\eta=0.4$, and the homeoidally stratified
axisymmetric Hernquist model [equation~(\ref{eqtrihern})] with
$\epsilon=0$ and $\eta=0.5$.}
\label{figcurl}
\end{figure}

\section{A simple application: triaxial Hernquist galaxy}
\label{secappl}

In order to illustrate the method described in Section~2 we now
construct a triaxial galaxy model with a realistic density
distribution and analytical MOND potential.

Following Step 1, we start with the spherically symmetric Hernquist
(1990) density distribution
\begin{equation}
\rhoz(r)={\Mz\over 2\pi\rc^3}{1 \over s(1+s)^3},
\end{equation}
where $s\equiv r/\rc$; we recall that this density distribution, when
projected, is a good approximation of the $R^{1/4}$ law (de
Vaucouleurs 1948) over a large radial interval, thus providing a
realistic description of the luminous density distribution of
elliptical galaxies.  From equation~(\ref{eqgz})
$\gz=\sqrt{G\Mz\az}/\rc(1+s)$, and so the dMOND potential is
\begin{equation}
\phiz =\sqrt{G\Mz\az}\ln(1+s),
\label{eqphizher}
\end{equation}
while from equation~(\ref{eqrhoNz}) 
\begin{equation}
\rhoNz ={\sqrt{G\Mz\az}\over 4 \pi G \rc^2} {s+2\over s(s+1)^2}.
\end{equation}
In agreement with the discussion in Section~2, $\rhoNz\sim\rhoz\sim
r^{-1}$ in the central regions and $\rhoNz \propto r^{-2}$ in the
outer parts of the dMOND Hernquist model .

According to Step 2, we look for a density distribution $\rhoNu$ to be
added to $\rhoNz$ such that the total density is positive and the
potential $\phiu$ is explicitly known.  A family of densities that
can be easily expanded with the CB05 technique and admit a simple
analytical potential $\phiu$ is
\begin{equation}
\varrho(r)={\varrhoz\beta^a\over (s+\beta)^a},\quad (a>3),
\label{eqvarrho}
\end{equation}
where $\varrhoz$ is a density scale, and $\beta$ is the scale-length
in units of $\rc$.  There is nothing special about the distribution
(\ref{eqvarrho}), which has been chosen only by simplicity arguments:
note however that at both small and large radii the total density is
dominated by $\rhoNz$, so its positivity has to be checked only for
the intermediate regions.  Here we restrict to the case $a=6$ (see
also Muccione \& Ciotti~2004), and from equation~(\ref{eqrhoNu})
\begin{equation}
\rhoNu=-6\varrhoz\beta^6{(\epsilon\sthsq\sphsq+\eta\cthsq)s \over (s+\beta)^7},
\end{equation}
so that
\begin{equation}
\phiu =4\pi G\varrhoz\beta^5\rc^2\left[ 
        {(\epsilon+\eta)(s^2+3 \beta s+\beta^2)\over 30\beta^2(s+\beta)^3}+
        {(\epsilon\sthsq\sphsq+\eta\cthsq)s^2\over 5(s+\beta)^5}
                    \right].
\label{eqphiusix}
\end{equation}

Before computing the dMOND operator for $\phiz+\phiu$ it is worth
studying its linearization for $\epsilon\to0$ and $\eta\to0$, and from
equation~(\ref{eqlinrho}) we obtain
\begin{eqnarray}
{\rho \rc^3 \over \Mz} & \simeq & {1 \over 2\pi s (s+1)^3} + \beta^3(\tilde{\epsilon}+\etatilde)
{s^5+7\beta s^4+9\beta^2s^3+\beta^2(\beta-6)s^2-2\beta^3(\beta+6)s-6\beta^4
\over
15(s+1)^2(s+\beta)^7}
+\nonumber\\
     &  & 6\beta^5(\epsilontilde\sthsq\sphsq+\etatilde\cthsq)
{6 s^3-(23\beta-3)s^2+\beta(\beta-24)s+3\beta^2 
\over
15(s+1)^2(s+\beta)^7},
\label{eqlinher}
\end{eqnarray} 
where the new dimensionless parameters $\epsilontilde$ and $\etatilde$
are defined as
$\epsilontilde/\epsilon=\etatilde/\eta=\varrhoz\rc^2\sqrt{G/\Mz\az}$.
Thus, $\rho \sim \rhoz \sim r^{-1}$ for $r \to 0$, while for $r \to
\infty$
 \begin{eqnarray}
{\rho\rc^3 \over \Mz} \simeq \left({1 \over 2\pi} +
\beta^3{\epsilontilde+\etatilde\over 15}\right){1\over s^4}+ 12\beta^5
{\epsilontilde\sthsq\sphsq+\etatilde\cthsq \over 5s^6},
\label{eqasynt}
\end{eqnarray} 
i.e., the linearized density retains the spherical symmetry and radial
trend of $\rhoz$ for $r\to 0$ and $r \to \infty$, so negative values
of $\rho$ can be present at intermediate distances only from the
center. For instance, restricting to the axisymmetric case
($\epsilon=0$) with $\beta=5$, according to
equation~(\ref{eqlinher})negative values of $\rho$ first appear for
$\etatilde\gsim0.018$ along the $z$-axis near $z\sim4\rc$, and smaller
values of $\etatilde$ correspond to an everywhere positive $\rho$.

The formula resulting from the evaluation of the dMOND operator is not
reported here, and we limit ourselves to show the isodensity contours
in a pair of representative axisymmetric ($\epsilon=0$) cases, fixing
$\beta=5$.  In Fig.~\ref{figden} (top left panel) we show the
isodensity contours in the meridional plane for the model with
$\etatilde=0.01$. Note how the density is spherically symmetric both
in the central regions and at large distances, in accordance with
equation~(\ref{eqlinher}). It is particularly important to stress that
the maximum flattening of the density corresponds to an axis ratio
$\sim 0.8$, even though $\etatilde$ is much smaller than 0.2. This at
variance with the Newtonian case (in which the flattening of the
expanded density is the same as that of the seed density for small
deformations), and this is a result of the non-linear nature of
MOND. In the top right panel of Fig.~\ref{figden} we show the density
profiles along the symmetry axis (solid line) and along a radius in
the equatorial plane (dashed line) for the same model.  It is apparent
how the radial trend of the spherical Hernquist model has been nicely
preserved by the imposed deformation, again in accordance with
equation~(\ref{eqlinher}). Note, however, how the density profile
along the $z$ axis shows a small depression, which becomes more
apparent in the model $\etatilde=0.02$ (Fig.~\ref{figden}, lower
panels).  This last model is near the consistency limit in
$\etatilde$, and the axis ratio of the flattest isodensity surface is
$\sim 0.6$.  Larger values of $\etatilde$ would produce a negative
density along the $z$ axis at $z\sim4\rc$, in agreement with the
results of the linear analysis.

It should be clear that the presented model is just one of the many
analytical triaxial Hernquist models one can construct. For example,
one could adopt the $\phiu$ derived from the homeoidal expansion of a
power-law $\varrho$ (CB05), or just add a quadrupole potential [i.e.,
assume $\psiu=0$ in equation~(\ref{eqphiu})]. For instance, in
Fig.~\ref{figdenbis} we show the density obtained by evaluating the
dMOND density of the potential $\phiz+\phiu$, where $\phiz$ is still
given by equation~(\ref{eqphizher}), but
\begin{equation} 
\phiu=\sqrt{G\Mz\az} {(\epsilon\sthsq\sphsq+\eta\cthsq)s\over
  (s+\beta)^2}
\label{eqphiuq}
\end{equation} 
The two models shown are characterized by $\epsilon=0$, $\beta=1$,
$\eta=0.2$ (top) and $\eta=0.4$ (bottom). Overall, we can note that
for these models the isodensity surfaces are flat also in the central
regions and in the models outskirts, and that the symmetry axis is
again the region where the model becomes unphysical for too large
$\eta$.

We conclude this Section discussing briefly the global properties of
the dMOND field $\Sv=g\gv/\az-\gvN$ for the presented models. In
Fig.~4 (left) we show the field $\Sv/\gN$ in the meridional plane,
while in the right panels we plot lines of constant $S/\gN$, where
$S\equiv||\Sv||$ (The Newtonian field $\gvN$ has been evaluated
numerically with a spectral Poisson solver; see Section~4).  The
quantity $S/\gN$ gives direct information about the relative
importance of the solenoidal field with respect to the Newtonian
gravitational field. Remarkably, as already pointed out by BM95, also
in our case the solenoidal field is typically almost everywhere much
smaller than $\gN$: we find $S/\gN\lsim0.05$ for the family of
axisymmetric and triaxial models obtained from
equation~(\ref{eqphiusix}). However, we also find that this is not
necessarily the rule: for example, $S/\gN\sim0.25$ in the central
regions of the models obtained from equation~(\ref{eqphiuq}), and
other cases of non-negligible solenoidal field will be discussed in
the next Section.

\section{The numerical MOND potential solver}
\label{msolv}

The results presented in the previous Sections, albeit encouraging,
suggest that in order to have a full control on the density field
producing the MOND potential the best tool is still the numerical
solution of equation~(\ref{eqMOND}). To our knowledge the only MOND
numerical solvers presently available are the two-dimensional code
developed by Milgrom~(1986), and the BM95 three-dimensional,
multi-grid solver [used by Brada \& Milgrom (1999) in their N-body
code].  Here, as a complement to the analytical method, we present a
new numerical three-dimensional MOND potential solver based on a
spherical coordinates grid, which can be easily implemented in a
standard particle-mesh N-body code.

The goal is to solve numerically equation~(\ref{eqMOND}) for
$\phi(\xx)$, i.e. the non--linear elliptic boundary value problem
defined by
\begin{equation}
\hM[\phi(\xx)]\equiv\nabla\cdot\left[\mu\left({g \over
    \az}\right)\nabla\phi(\xx)\right]-4\pi G\rho(\xx)=0,\quad 
    g=O(r^{-1})\,\,{\rm for }\,\,r\to\infty,
\label{eqdiv}
\end{equation}
where $\xx=(x,y,z)$, $r\equiv\Vert \xx \Vert$, $\rho(\xx)$ is
assigned, and $g=\Vert\gv\Vert$ with $\gv=-\nabla\phi(\xx)$.

The numerical scheme is based on the iterative Newton method, which is
robust and relatively simple to implement. At the $(n+1)-th$ iteration
we indicate with $\dphinow$ the increment function to be determined so
that $\phiplus=\phinow+\dphinow$ is a better approximation of the
solution, while $\phinow$ and $\gnow=-\nabla\phinow$ are quantities
known from the previous iteration. In our case, $\dphinow$ is the
solution of the linear equation 
\begin{equation}
\dhMnow\left[\dphinow\right]=-\hM\left[\phinow\right],
\label{eqnew}
\end{equation} 
where the linear operator 
\begin{equation}
\dhMnow \equiv \nabla\cdot\left[ \munow\nabla+ \muprimenow \gvnow
{\left(\gvnow\cdot\nabla\right)} \right]
\label{eqdhM}
\end{equation}
satisfies the identity
\begin{equation}
\hM\left[\phiplus\right]-\hM\left[\phinow\right]=\dhMnow\left[\dphinow\right]+O\left[(\dphinow)^2\right], 
\label{eqhMplus}
\end{equation}
where $\munow\equiv\mu(\gnow/\az)$ and $\muprimenow\equiv \mu^{\prime}
\left({\gnow/\az}\right)/\gnow\az$ are known. Equation~(\ref{eqnew})
is solved by inversion of $\dhMnow$, and the procedure is repeated
until numerical convergence to the solution $\phinum$ (defined by
$\max\vert \hM[\phinum] \vert < \varepsilon$, where $\varepsilon$ is a
prescribed tolerance) is attained.

The solution of equation~(\ref{eqnew}) requires in general a finite
difference approximation of the spatial derivatives and the inversion
of a three-dimensional matrix.  Note that boundedness of the inverse
of the operator $\dhMnow$ assures quadratic convergence of the scheme
for $\phi^{(0)}$ sufficiently close to the sought solution (e.g. Stoer
\& Bulirsch~1980).  In our implementation, designed for finite mass
distributions, we represent equation~(\ref{eqdiv}) and (\ref{eqnew})
in spherical coordinates, which allow for a simple assignment of the
boundary conditions. Moreover, we approximate the linear
operator~(\ref{eqdhM}) with
\begin{equation} 
\dhMbar\equiv{1\over r^2}\left[{\partial \over \partial
r}\left({r^2\mubar(r){\partial \over \partial r}}\right) 
+\mubar(r)\left(\hL_{\vartheta}+\hL_{\varphi}\right)\right],
\label{eqLap}
\end{equation}
where
\begin{equation}
\hL_{\vartheta}\equiv{1\over\sth}{\partial\over\partial\vartheta}\left(
\sth{\partial\over\partial\vartheta}\right),\quad
\hL_{\varphi}\equiv{1\over\sth}
{\partial^2\over\partial\varphi^2}
\end{equation}
are the angular components of the Laplace operator,
$\mubar(r)=(1/4\pi)\int \munow(r,\vartheta,\varphi) \sth d\vartheta
d\varphi$, and the variation $\muplus-\munow \propto \muprimenow$ is
neglected.  It is easy to show that with this new operator quadratic
convergence in equation~(\ref{eqnew}) is replaced by linear
convergence since
\begin{equation}
\hM\left[\phiplus\right]-\hM\left[\phinow\right]=\dhMbar\left[\dphinow\right]+O\left[\dphinow\right].
\end{equation}
However, the inversion of $\dhMbar$ is much simpler because we are now
in the position to use the full power of spherical harmonics. In fact,
after expanding the source term $\hM\left[\phinow\right]$ and the
unknown function $\dphinow$ in spherical harmonics
\begin{equation}
\dphinow(r,\vartheta,\varphi)=\sum_{l,m}\dphinowlm(r)Y_l^m(\vartheta,\varphi)
\label{eqlegendre}
\end{equation}
(e.g. Binney \& Tremaine~1987), the modified equation~(\ref{eqnew})
reduces to
\begin{equation}
{1\over r^2}\left[{\partial \over \partial r}\left({r^2\mubar{\partial
      \over \partial r}}\right)
-\mubar(r)l(l+1)\right]\dphinowlm(r)=-\hM\left[\phinow \right]_{l,m}, 
\label{eqdphilm}
\end{equation}
involving derivatives only in the radial coordinates.  The solutions
$\dphinowlm(r)$ are back transformed into
$\dphinow(r,\vartheta,\varphi)$ according to
equation~(\ref{eqlegendre}), and the new components of the
acceleration field $\gvplus=\gvnow-\nabla\dphinow$ are evaluated. 

In order to solve equation~(\ref{eqdphilm}) we introduce the
invertible mapping for the radial coordinate
\begin{equation}
r(\xi)=\tan^{\alpha}\xi,\quad  (0\le\xi < \pi/2),
\label{eqmapping}
\end{equation}
so that the infinite radial range is mapped onto the finite interval
$[0,\pi/2)$, 
\begin{equation}
{\partial\over\partial r}={\cos^2\xi\over \alpha\tan^{\alpha-1}\xi}{\partial\over\partial\xi},
\end{equation}
and the boundary conditions for the radial derivative of the potential
at $\xi=\pi/2$ are automatically satisfied (Londrillo \&
Messina~1990); in our applications we adopt $\alpha=1$ or $\alpha=2$.
In practice, the $(\xi,\vartheta,\varphi)$ coordinates are discretized
on a uniform $(\xi_i,\vartheta_j,\varphi_k)$ grid of
$\Nxi\times\Nth\times\Nph$ points ($\Nxi=64$, $\Nth=32$, and
$\Nph=64$, in typical applications), the $\vartheta$ and $\varphi$
derivatives are evaluated using the spectral representation of finite
order Legendre-Fourier polynomials, and the radial derivatives are
approximated by second order centered differences in the $\xi_i$
coordinate.  Thus, for each $(l,m)$ the discretized operator
$\dhMbar_{l,m}$ to be inverted is represented by a pentadiagonal
matrix over $\xi_i$. We note that for $l>0$ we solve
equation~(\ref{eqdphilm}) for $\dphi^{(n)}_{l,m}$ with the boundary
condition $\dphi^{(n)}_{l,m}(\xi=\pi/2)=0$, while for $l=0$
equation~(\ref{eqdphilm}) can be solved directly for the radial
component of the acceleration increment.

\begin{figure}[h]
\epsscale{.50}
\plotone{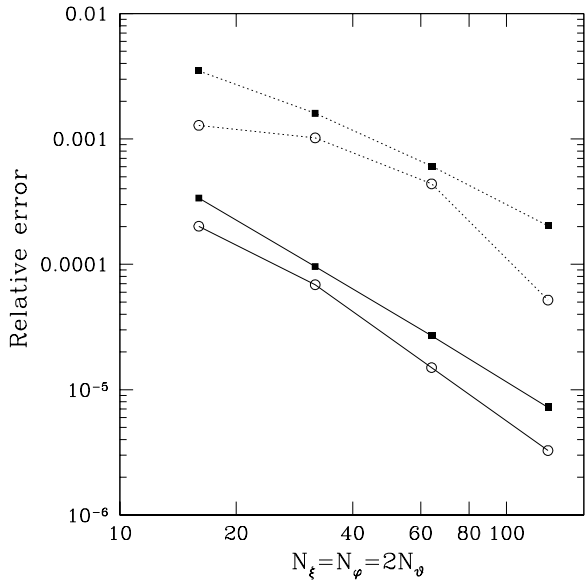}
\caption{R.m.s. (solid line) and maximum (dotted line) relative error
  on the acceleration (calculated over all the space) as a function of
  the number of radial grid points $\Nxi=\Nph=2\Nth$ for two triaxial
  Hernquist models. Empty circles refer to the model with $\phiu$ in
  equation~(\ref{eqphiusix}) with $\beta=5$, $\etatilde=0.02$ and
  $\epsilontilde=0.01$, while solid squares to the model with $\phiu$
  in equation~(\ref{eqphiuq}) with $\beta=1$, $\eta=0.4$ and
  $\epsilon=0.2$.}
\label{figerr}
\end{figure}

\subsection{The tests}
\label{test}

We tested the numerical code using the analytical axisymmetric and
triaxial MOND models described in Section~3.  We assigned the
analytical density distribution obtained for different values of
$\az\rc^2/(G\Mz)$, $\epsilon$, $\eta$, and $\beta$, and we recovered
the corresponding numerical MOND potential, comparing it with its
analytical expression.  The accuracy of the numerically estimated
field $\gvnum\equiv-\nabla \phinum$ is quantified by the relative
error $||\gv-\gvnum||/g$, and in all the studied cases we found very
good agreement between the analytical and numerical acceleration
field, even for models near the consistency limit such as those in the
bottom panels of Fig.~\ref{figden} and \ref{figdenbis}.  For example,
in Fig.~\ref{figerr} we plot the r.m.s. (solid line) and maximum
(dotted line) relative error calculated over all grid points, as a
function of $\Nxi=\Nph=2\Nth$ for two triaxial Hernquist models: one
with $\phiu$ as in equation~(\ref{eqphiusix}) (empty symbols), the
other with $\phiu$ as in equation~(\ref{eqphiuq}) (solid symbols).  It
is apparent how the typical errors decrease down to values
$\sim10^{-5}$; such values are reached in $15-20$ iterations, taking
few seconds on a common PC. 

As a second set of tests, we studied the razor-thin Kuzmin disk
(BM95). This is a quite severe test, because of the singular nature of
the density distribution along the $z$ direction. In this case, we
found numerical convergence to the solution, with relative
r.m.s. error $5.6\times10^{-3}$ for $\Nxi=\Nph=128$, and $\Nth=64$.
 
We note that our code converges also when dealing with multi-centered
density distributions, for which the spherical grid is clearly not
optimal.  In such cases the convergence is not as fast as in the case
of density distributions centered in the origin. For example, the MOND
fields of groups of 2, 3, and 4 Hernquist spheres are computed in
$20-25$ iterations, with tolerance $\varepsilon\sim 10^{-5}$,
$\Nxi=64$, $\Nth=32$, and $\Nph=64$.

As an illustrative application of the code, we evaluated the dMOND
field for the family of homeoidally stratified triaxial Hernquist
models
\begin{equation}
\rho= {\Mz \over 2\pi \rc^3(1-\epsilon)(1-\eta) m (1+m)^3},
\end{equation}
where
\begin{equation}
m^2={x^2\over\rc^2}+{y^2\over\rc^2(1-\epsilon)^2}+{z^2\over\rc^2(1-\eta)^2},
\label{eqtrihern}
\end{equation}
and we investigated their solenoidal field for different flattenings.
For example, the model in Fig.~{\ref{figcurl}} (bottom panels) has
$\epsilon=0$ and $\eta=0.5$.  The global behavior of $\Sv$ is similar
to what found for the analytical models of Section~3, with the larger
contribution of $\Sv$ near the center. In this case we found maximum
values of $S/\gN\sim 0.3$. Even larger values were found for more
flattened models: for instance, $0.2\lsim S/\gN \lsim0.6$ in the
central regions ($r\lsim\rc$) of a triaxial model with $\eta=0.8$,
$\epsilon=0.6$.  This is curious because BM95 found similar values of
$S/\gN$ only in the quite artificial case of disks with a central
hole. This last result suggests that when studying MOND equilibrium
models for disks or elliptical galaxies it could be dangerous to solve
equation~(\ref{eqmug}) just relying on the {\it Ansatz} that in any
case $\Sv$ would be small.

\section{Summary and conclusions}

In this paper we presented a few representative axisymmetric and
triaxial density models, both analytical and numerical, obeying the
MOND field equation. The analytical density-potential pairs are
constructed by means of a simple method based on the deformation of
the potential of spherically symmetric systems. We show that in this
way it is possible to build systems with radial profiles and shapes
similar to those of real galaxies. Our method, although applied in the
present paper by using simple deformations is easily generalizable to
more complicated cases.  As a complement to the analytical method, we
also presented a numerical code (based on spectral methods) to solve
the MOND field equation. We tested the code against the new analytical
triaxial models and the MOND Kuzmin disk, with excellent results in
terms of both computational time and accuracy of the numerical
solution.  As a first simple application of the code, we explored the
relevance of the solenoidal field $\Sv$ in MOND distributions. Though
we confirm that this field is typically small compared to the
Newtonian field of the density distribution, we also found that for
some systems $\Sv$ is certainly not negligible, at least in some
regions of space.

A particularly important application of the numerical solver will be
its implementation in a particle-mesh N-body code. Such an
implementation is promising, because our MOND solver is based on an
iterative scheme, and at each time in a N-body simulation the
potential at the previous time is a very good seed for the iterative
procedure that should converge efficiently.  More immediate possible
applications of the presented analytical models and numerical code are
the study of MOND orbits in aspherical density distributions, and the
vertical motions of stars near the galactic plane.  We conclude by
pointing out that the developed numerical code can also be used to
solve the equation for the scalar field in the non-relativistic limit
of TeVeS, the relativistic MOND theory introduced by Bekenstein~(2004;
equation~53), allowing, for example, to investigate MOND gravitational
lensing from non-spherical lenses.


We are grateful to James~Binney, Hongsheng~Zhao, and the anonymous
referee for very useful comments.  This work was partially supported
by a MIUR grant CoFin2004.

\appendix

\section{Three existence theorems}

We present here three existence theorems that are at the basis of the
technique presented in Section 2. The first result holds in general,
while the second and third results assume spherical symmetry.

{\bf Theorem 1} {\it Let $\phiN$ the Newtonian potential 
generated by a positive density distribution, i.e.
\begin{equation}
\nabla^2 \phiN\geq 0
\end{equation} 
over the whole space. If 
\begin{equation}
\nabla\cdot (||\nabla\phiN ||\nabla\phiN)\geq 0
\end{equation}
over the whole space, then also 
\begin{equation}
\nabla\cdot \left[\mu \left({||\nabla\phiN ||\over \az}\right)\nabla\phiN\right]\geq 0,
\end{equation}
where $\mu$ is given in equation~(\ref{eqmu}).
}

{\bf Proof} 
Equation~(A3) can be rewritten as
\begin{equation}
\nabla\cdot \left[\mu \left({\gN\over \az}\right)\nabla\phiN\right]=
{1\over\sqrt{\az^2+\gN^2}}
\left (\gN \nabla^2\phiN + 
{\nabla\gN\cdot\nabla\phiN \over 1+\gN^2/\az^2}\right ),
\end{equation}
from condition (A1) a negative density can be produced only in the
region of space $\Omega^-$ where $\nabla \gN\cdot\nabla\phiN <0$. If
$\Omega^-=\emptyset$ the theorem is proved, thus let $\Omega^-\neq
\emptyset$.  In this case, by expansion of inequality (A2) we have
that on $\Omega^-$
\begin{equation}
\gN\nabla^2\phiN\;\geq\; -\nabla \gN \cdot \nabla\phiN 
\;\geq\;-
{\nabla \gN \cdot \nabla\phiN \over 1+\gN^2/\az^2},
\end{equation} 
and this proves the theorem.

We now prove a consequence of Theorem 1 holding for spherically
symmetric systems, which is relevant to the construction of exact
aspherical MOND solutions.

{\bf Theorem 2} {\it Let $\phi (r)$ the dMOND potential of a
spherically symmetric (positive) density distribution. Then
\begin{equation}
\nabla^2 \phi\geq 0,
\label{eqlaplpos}
\end{equation} 
and 
\begin{equation}
\nabla\cdot \left[\mu \left({||\nabla\phi ||\over \az}\right)\nabla\phi\right]\geq 0,
\label{eqMONDpos}
\end{equation}
over the whole space.
}

{\bf Proof} Direct calculation of $\nabla^2 \phi$ in spherical
symmetry, as done in equation~(\ref{eqrhoNz}), proves
equation~(\ref{eqlaplpos}). Conditions (A1) and (A2) are then verified
and this proves equation~(\ref{eqMONDpos}). 

Note that the converse of Theorem 2 is not true, i.e.,
$\nabla^2\phi\geq 0$ is not a sufficient condition for
$\nabla\cdot(||\nabla\phi ||\nabla\phi)\geq 0$. In fact, the following
result holds

{\bf Theorem 3} {\it Let 
\begin{equation}
\nabla^2 \phi=4\pi G\rho(r)\geq 0.
\end{equation} 
Then 
\begin{equation}
\nabla\cdot (||\nabla\phi ||\nabla\phi)\geq 0,
\label{eqdmondpos}
\end{equation}
if and only if
\begin{equation}
\rho(r)\geq {M(r)\over 4\pi r^3}\quad\forall r.
\end{equation}
}

{\bf Proof} The proof is obtained by expanding
inequality~(\ref{eqdmondpos}) in spherical symmetry as
\begin{equation}
{d\phi\over dr}\left({d^2\phi\over dr^2}+4\pi G\rho\right)\geq 0:
\end{equation}
the identity $d\phi/dr=GM(r)/r^2$ proves the theorem.

\end{document}